\documentclass[epj,final,nopacs]{myownsvjour}
\usepackage{bm,color,graphicx}
\newcommand{\beq}{\begin{equation}}
\newcommand{\eeq}{\end{equation}}
\newcommand{\beqa}{\begin{eqnarray}}
\newcommand{\eeqa}{\end{eqnarray}}

\newcommand{\abs}[1]{\left\vert#1\right\vert}
\newcommand{\dd}{{\rm d}}
\newcommand{\der}[1]{\frac{\dd#1}{\dd t}}
\newcommand{\diag}{\mathop{{\rm diag}}}
\newcommand{\e}{{\rm e}}
\newcommand{\frad}[2]{\displaystyle{\displaystyle#1\over\displaystyle#2}}
\newcommand{\hi}{{\rm max}}
\newcommand{\lo}{{\rm min}}
\newcommand{\ii}{{\rm i}}
\newcommand{\m}[1]{{\bm#1}}
\newcommand{\mean}[1]{\left\langle#1\right\rangle}
\newcommand{\muh}{{\frad{\mu}{2}}}
\newcommand{\qh}{{\frad{q}{2}}}
\newcommand{\prob}[1]{\mathop{{\rm Prob}}\left\{#1\right\}}
\newcommand{\ze}{{(0)}}
\newcommand{\un}{{(1)}}
\newcommand{\de}{{(2)}}

\newcommand{\Lim}{{\lim_{N\to\infty}}}
\renewcommand{\P}{{\cal P}}
\newcommand{\Q}{{\cal Q}}

\begin{document}

\title{On the coexistence of competing languages}

\author{Jean-Marc Luck\inst{1} \and Anita Mehta\inst{2}}

\institute{
Universit\'e Paris-Saclay, CNRS, CEA, Institut de Physique Th\'eorique,
91191 Gif-sur-Yvette, France.\\
\email{jean-marc.luck@ipht.fr}
\and
Centre for Linguistics and Philology, University of Oxford,
Walton Street, Oxford OX1 2HG, UK.\\
\email{anita.mehta@ling-phil.ox.ac.uk}}

\date{}

\abstract{
We investigate the evolution of competing languages,
a subject where much previous literature suggests that the outcome
is always the domination of one language over all the others.
Since coexistence of languages is observed in reality,
we here revisit the question of language competition,
with an emphasis on uncovering the ways in which coexistence might emerge.
We find that this emergence is related to symmetry breaking,
and explore two particular scenarios
-- the first relating to an imbalance in the population dynamics
of language speakers in a single geographical area,
and the second to do with spatial heterogeneity,
where language preferences are specific to different geographical regions.
For each of these,
the investigation of paradigmatic situations
leads us to a quantitative understanding
of the conditions leading to language coexistence.
We also obtain predictions of the number of surviving languages
as a function of various model parameters.}

\maketitle

\section{Introduction}
\label{intro}

The dynamics of language evolution is one of many interdisciplinary fields
to which methods and insights from statistical physics have been successfully applied
(see~\cite{cfl} for an overview, and~\cite{bly} for a specific comprehensive review).

In this work we revisit the question of language coexistence.
It is known that a sizeable fraction of the more than 6000 languages
that are currently spoken, is in danger of becoming
extinct~\cite{krauss,crystal,sutherland}.
In pioneering work by Abrams and Strogatz~\cite{as}, theoretical predictions
were made to the effect
that less attractive or otherwise unfavoured languages are generally doomed to extinction,
when contacts between speakers of different languages become sufficiently frequent.
Various subsequent investigations have corroborated this finding,
emphasising that the simultaneous coexistence of competing languages
is only possible in specific circumstances~\cite{sce,zg},
all of which share the common feature that
they involve some symmetry breaking mechanism~\cite{bly}.
A first scenario can be referred to as spatial symmetry breaking.
Different competing languages may coexist in different geographical areas,
because they are more or less favoured locally,
despite the homogenising effects of migration and language
shift~\cite{pl,ces,fam}.
A second scenario corresponds to a more abstract internal symmetry breaking.
Two or more competing languages may coexist at a given place
if the populations of speakers of these languages have
imbalanced dynamics~\cite{pr,ks,kus}.
Moreover, it has been shown that a stable population of bilinguals
or multilinguals also favours the coexistence of several
languages~\cite{mp,mw,cmm}.

The aim of the present study is to provide a quantitative understanding
of the conditions which ensure the coexistence of two or more competing languages
within each of the symmetry breaking scenarios outlined above.
Throughout this paper, in line with many earlier
studies on the dynamics of languages~\cite{as,zg,pl,fam,pr,ks,kus,mp,mw,cmm},
and with an investigation of grammar acquisition~\cite{knn},
we describe the dynamics of the numbers of speakers of various languages
by means of coupled rate equations.
This approach is sometimes referred to as ecological modelling,
because of its similarity with models used in theoretical ecology (see e.g.~\cite{berryman}).
From a broader perspective,
systems of coupled differential equations,
and especially Lotka-Volterra equations and replicator equations,
are ubiquitous in game theory and in a broad range of areas in mathematical biology
(see e.g.~\cite{hofbook,hofbams,nsscience}).

The plan of this paper is as follows.
For greater clarity, we first consider in Section~\ref{internal}
the situation of several competing languages in a single geographic area where
the population is well mixed.
We address the situation where internal symmetry is
broken by imbalanced population dynamics.
The relevant concepts are reviewed in detail in the case of two competing
languages in Section~\ref{itwo}, and the full phase diagram of the model is derived.
The case of an arbitrary number $N$ of competing languages
is then considered in Section~\ref{in} in full generality.
The special situation where the attractivenesses of the languages are equally
spaced is studied in Section~\ref{iorder},
whereas Section~\ref{igal} is devoted to the case
where attractivenesses are modelled as random variables.
Section~\ref{spatial} is devoted to the
situation where coexistence is due to spatial symmetry breaking.
We focus our attention onto the simple case of two languages in competition
on a linear array of $M$ distinct geographic areas.
Language attractivenesses vary arbitrarily along the array,
whereas migrations take place only between neighbouring areas
at a uniform rate~$\gamma$.
A uniform consensus is reached at high migration rate,
where the same language survives everywhere.
This general result is demonstrated in detail for two geographic areas
(Section~\ref{stwo}),
and generalised to an arbitrary number $M$ of areas (Section~\ref{sm}).
The cases of ordered and random attractiveness profiles
are investigated in Sections~\ref{sorder} and~\ref{srandom}.
In Section~\ref{disc} we present a non-technical discussion of our findings and their implications.
Two appendices contain technical details
about the regime of a large number of competing languages in a single
geographic area (Appendix~\ref{appa})
and about stability matrices and their spectra (Appendix~\ref{appb}).

\section{Breaking internal symmetry:
language coexistence by imbalanced population dynamics}
\label{internal}

This section is devoted to the dynamics of languages in a single geographic area.
As mentioned above, it has been shown that two or more competing languages may coexist
only if the populations of speakers of these languages have
imbalanced dynamics~\cite{pr,ks,kus}.
Our goal is to make these conditions more explicit
and to provide a quantitative understanding of them.

\subsection{Two competing languages}
\label{itwo}

We begin with the case of two competing languages.
We assume that language~1 is more favoured than language~2.
Throughout this work we neglect the effect of bilingualism,
so that at any given time $t$
each individual speaks a single well-defined language.
Let $X_1(t)$ and $X_2(t)$ denote the numbers of speakers of each language at
time $t$,
so that $X(t)=X_1(t)+X_2(t)$ is the total population of the area under
consideration.

The dynamics of the model is defined by the coupled rate equations
\beqa
\der{X_1(t)}=X_1(t)(\,\underbrace{1-X_1(t)-qX_2(t)}+CX_2(t)),
\label{x1}
\\
\der{X_2(t)}=X_2(t)(\,\underbrace{1-X_2(t)-qX_1(t)}-CX_1(t)).
\label{x2}
\eeqa

The above equations are an example of Lotka-Vol\-ter\-ra equations
(see e.g.~\cite{berryman,hofbook}).
The terms underlined by braces
describe the intrinsic dynamics of the numbers of speakers of each language.
For the sake of simplicity
we have chosen the well-known linear-minus-bilinear or `logistic' form
which dates back to Lotka~\cite{lotka}
and is still commonly used in population dynamics.
The linear term describes population growth,
whereas the quadratic terms represent a saturation mechanism.

The main novelty of our approach
is the introduction of the parameter~$q$ in the saturation terms.
This imbalance parameter
is responsible for the internal symmetry breaking leading to language coexistence.
It allows for the interpolation between two situations:
when the saturation mechanism only involves the total population, i.e., $q=1$,
and when the saturation mechanism acts separately on the populations of speakers
of each language, $q=0$, which is
the situation considered by Pinasco and Romanelli~\cite{pr}.
Generic values of $q$ correspond to tunably imbalanced dynamics.

The last term in each of equations~(\ref{x1}),~(\ref{x2})
describes the language shift consisting of the conversions of single
individuals
from the less favoured language~2 to the more favoured language~1.
In line with earlier studies~\cite{zg,pr,ks,kus},
conversions are triggered by binary interactions between individuals,
so that the frequency of conversions is proportional to the product $X_1(t)X_2(t)$.
The reduced conversion rate $C$ measures the difference of attractivenesses
between the two languages.

For generic values of the parameters $q$ and $C$,
the rate equations~(\ref{x1}),~(\ref{x2}) admit a unique stable fixed point.
The dynamics converges exponentially fast to the corresponding stationary state,
irrespective of initial conditions.
There are two possible kinds of stationary states:

\begin{itemize}

\item
{\it I.~Consensus.}

The solution
\beq
X_1=1,\quad X_2=0,\quad X=1
\eeq
describes a consensus state where the unfavoured language~2 is extinct.
The inverse relaxation times describing convergence toward the latter state
are the opposites of the eigenvalues of the stability matrix
associated with equations~(\ref{x1}),~(\ref{x2}).
The reader is referred to Appendix~\ref{agal} for details.
These inverse relaxation times read
\beq
\omega_1=1,\quad\omega_2=q+C-1.
\label{Irelax}
\eeq
The above stationary solution is thus stable whenever $q+C>1$.

\item
{\it II.~Coexistence.}

The solution
\beqa
X_1&=&\frac{1-q+C}{1-q^2+C^2},\quad
X_2=\frac{1-q-C}{1-q^2+C^2},
\nonumber\\
X&=&\frac{2(1-q)}{1-q^2+C^2}
\label{II}
\eeqa
describes a coexistence state where both languages survive forever.
This stationary solution exists whenever $q+C<1$.
It is always stable, as the inverse relaxation times read
\beq
\omega_1=1,\quad\omega_2=\frac{(1-q+C)(1-q-C)}{1-q^2+C^2}.
\label{IIrelax}
\eeq

\end{itemize}

\begin{figure}
\begin{center}
\includegraphics[angle=0,width=.7\linewidth,clip=true]{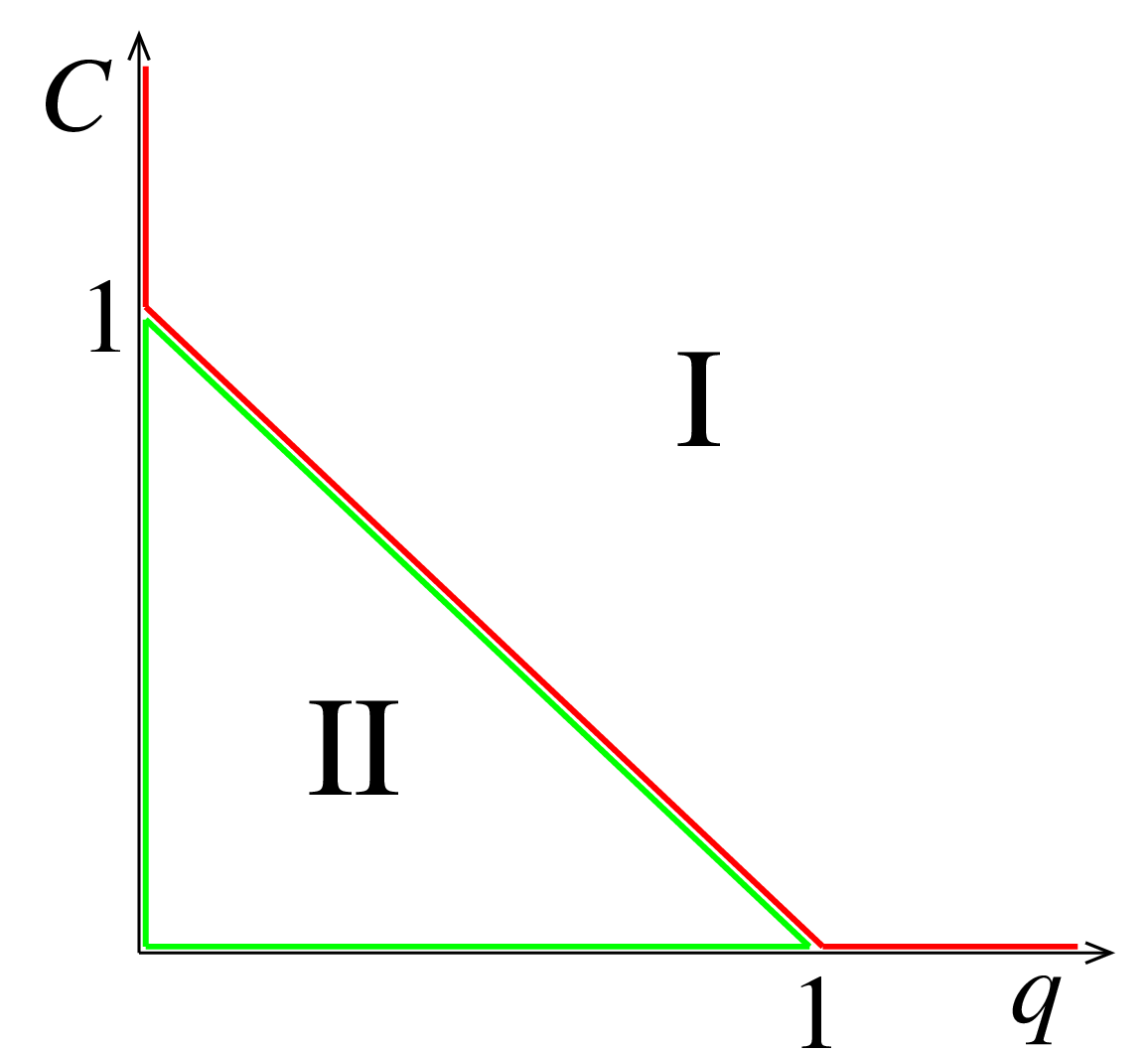}
\caption{
Phase diagram of the model in the $q$--$C$ plane.
I: consensus phase.
II: coexistence phase.}
\label{qcplot}
\end{center}
\end{figure}

Figure~\ref{qcplot} shows the phase diagram of the model in the~$q$--$C$ plane.
There is a possibility of language coexistence only for $q<1$.
The vertical axis ($q=0$) corresponds to the model considered by Pinasco and
Romanelli~\cite{pr},
where the coexistence phase is maximal and extends up to $C=1$.
As the parameter $q$ is increased,
the coexistence phase shrinks until it disappears at the point $q=1$,
corresponding to the balanced dynamics where the
saturation mechanism involves the total population.

The model exhibits a continuous transition
along the phase boundary between both phases ($q+C=1$).
The number $X_2$ of speakers of the unfavoured language
vanishes linearly as the phase boundary is approached from the coexistence
phase (see~(\ref{II})),
whereas the relaxation time $1/\omega_2$ diverges linearly as the phase
boundary is approached
from both sides (see~(\ref{Irelax}) and~(\ref{IIrelax})).

For parameters along the phase boundary ($q+C=1$),
the less attractive language still becomes extinct, albeit very slowly.
Equations~(\ref{x1}),~(\ref{x2}) here yield the power-law relaxation laws
\beqa
X_1(t)&\approx&1+\frac{2C-1}{2Ct},
\nonumber\\
X_2(t)&\approx&\frac{1}{2Ct},
\nonumber\\
X(t)&\approx&1+\frac{1}{t},
\eeqa
irrespective of initial conditions.

\subsection{$N$ competing languages}
\label{in}

The above setting
can be extended to the case of an arbitrary number $N$ of competing languages in a
given area.
Languages, numbered $i=1,\dots,N$, are more or less favou\-red,
depending on their attractivenesses~$A_i$.
The latter quantities are assumed to be quenched, i.e., fixed once for all.
This non-trivial static profile of attractivenesses is responsible for
conversions of single individuals
from less attractive to more attractive languages.

Let $X(t)$ be the total population of the area under consideration at time $t$,
and $X_i(t)$ be the number of speakers of language number $i=1,\dots,N$.
The dynamics of the model are defined by the rate equations
\beqa
\der{X_i(t)}=X_i(t)&\Biggl(&\underbrace{1-(1-q)X_i(t)-qX(t)}
\nonumber\\
&+&\sum_jC_{ji}X_j(t)\Biggr).
\label{xi}
\eeqa
The terms underlined by braces
describe the intrinsic dynamics of the numbers of speakers of each language.
The novel feature here is again the presence of the parameter~$q$,
which is responsible for imbalanced dynamics,
allowing thus the possibility of language coexistence.
The last term in~(\ref{xi}) describes the conversions of single individuals.
If language~$i$ is more attractive than language~$j$,
there is a net positive conversion rate $C_{ji}=-C_{ij}$ from language~$j$ to
language~$i$.
For the sake of simplicity,
we assume that these conversion rates depend linearly on the differences of
attractivenesses
between departure and target languages, i.e.,
\beq
C_{ji}=-C_{ij}=A_i-A_j,
\eeq
in some consistent units.

Throughout this work we shall not pay any attention to the evolution of the whole population $X(t)$.
We therefore reformulate the model in terms of the fractions
\beq
x_i(t)=\frac{X_i(t)}{X(t)}
\label{xdef}
\eeq
of speakers of the various languages,
which sum up to unity:
\beq
\sum_ix_i(t)=1.
\label{xsum}
\eeq

The reduction to be derived below is quite natural in the present setting.
It provides an example of the reduction of Lotka-Volterra equations
to replicator equations, proposed in ~\cite{hofna} (see also~\cite{hofbook,hofbams,nsscience}).
In the present situation, for $q<1$,
which is precisely the range of $q$
where there is a possibility of language coexistence,
the dynamics of the fractions~$x_i(t)$
obeys the following reduced rate equations,
which can be derived from~(\ref{xi}):
\beqa
\der{x_i(t)}&=&(1-q)X(t)\,x_i(t)
\nonumber\\
&\times&\Biggl(Z(t)-x_i(t)+\sum_jc_{ji}\,x_j(t)\Biggr),
\label{eqxi}
\eeqa
with
\beq
Z(t)=\sum_ix_i(t)^2,
\label{mudef}
\eeq
and where attractivenesses and conversion rates have been rescaled according to
\beqa
a_i&=&\frac{A_i}{1-q},
\\
c_{ji}&=&\frac{C_{ji}}{1-q}=a_i-a_j.
\label{acdefs}
\eeqa

In the following, we focus our attention onto the stationary states of the model,
rather than on its dynamics.
It is therefore legitimate to redefine time according to
\beq
t\to(1-q)\int_0^tX(t')\,\dd t',
\eeq
so that equations~(\ref{eqxi}) simplify to
\beq
\der{x_i(t)}=x_i(t)\Biggl(Z(t)-x_i(t)+\sum_jc_{ji}\,x_j(t)\Biggr).
\label{eq}
\eeq

The rate equations~(\ref{eq}) for the fractions of speakers of the $N$ competing languages
will be the starting point of further developments.
The quantity $Z(t)$ can be alternatively viewed as a dynamical Lagrange multiplier
ensuring that the dynamics conserves the sum rule~(\ref{xsum}).
The above equations belong to the class of replicator equations
(see e.g.~\cite{hofbook,hofbams,nsscience}).
Extensive studies of the dynamics of this class of equations have been made
in mathematical biology,
where the main focus has been on systematic classifications of fixed points and bifurcations
in low-dimensional cases~\cite{hofna,tj,bom1,s+s1,bom2,s+s2}.

From now on, we focus on the stationary state of the model
for arbitrarily high values of the number $N$ of competing languages.
The analysis of this goes as follows.
The stationary values $x_i$ of the fractions of speakers
are such that the right-hand sides of~(\ref{eq}) vanish.
For each language number~$i$, there are two possibilities:
either $x_i=0$, i.e., language $i$ gets extinct,
or $x_i>0$, i.e., language $i$ survives forever.
The non-zero fractions $x_i$ of speakers of surviving languages obey the
coupled linear equations
\beq
Z-x_i+\sum_j(a_i-a_j)x_j=0,
\label{eqfp}
\eeq
where the parameter $Z$ is determined by expressing
that the sum rule~(\ref{xsum}) holds in the stationary state.
For generic values of model parameters,
there is a unique stationary state,
and the system relaxes exponentially fast to the latter,
irrespective of its initial conditions.
The uniqueness of the attractor is characteristic of the specific form
of the rate equations~(\ref{eq}),~(\ref{eqfp}),
with skew-symmetric conversion rates $c_{ij}$ (see~(\ref{acdefs})).
This has been demonstrated explicitly
in the case of two competing languages, studied in detail in Section~\ref{itwo}.
The problem is however more subtle than it seems at first sight,
as the number~$K$ of surviving languages
depends on model parameters in a non-trivial way.

\subsection{The case of equally spaced attractivenesses}
\label{iorder}

It is useful to consider first the simple case
where the (reduced) attractivenesses~$a_i$ of the $N$ competing languages
are equally spaced between 0 and some maximal value that we denote by $2g$.
Numbering languages in order of decreasing attractivenesses,
so that language~1 is the most attractive
and language~$N$ the least attractive, this reads
\beq
a_i=g\,\frac{2N+1-2i}{N}.
\eeq
We have
\beq
\sum_ia_i=Ng.
\eeq
The parameter $g$ is therefore the mean attractiveness.

The (reduced) conversion rates read
\beq
c_{ji}=2g\,\frac{j-i}{N},
\eeq
so that the fixed-point equations~(\ref{eqfp}) take the form
\beq
Z-x_i+\frac{2g}{N}\sum_j(j-i)x_j=0.
\label{nfp}
\eeq
Already in this simple situation
the number~$K$ of surviving languages
depends on the mean attractiveness~$g$ in a non-trivial way.

Consider first the situation where all languages survive ($K=N$).
This is certainly true for $g=0$, where there are no conversions,
so that the solution is simply $x_i=1/N$.
There, all languages are indeed equally popular, as nothing distinguishes them.
More generally, as long as all languages survive,
the stationary solution obeying~(\ref{nfp}) reads
\beq
x_i=\frac{1}{N}+g\,\frac{N+1-2i}{N}=\frac{1}{N}-g+a_i
\label{xin}
\eeq
for $i=1,\dots,N$.
The above solution ceases to hold
when the fraction of speakers of the least attractive language vanishes, i.e., $x_N=0$.
This first extinction takes place for the threshold value
\beq
g_{N,N}=\frac{1}{N-1}
\eeq
of the mean attractiveness $g$.

\begin{figure}
\begin{center}
\includegraphics[angle=0,width=.7\linewidth,clip=true]{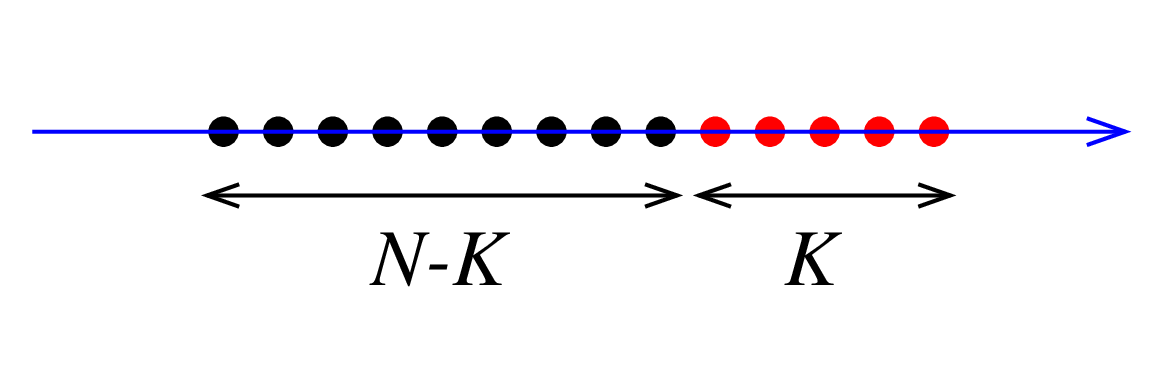}
\caption{
Sketch of the attractiveness axis.
Red symbols: $K$ surviving languages.
Black symbols: $N-K$ extinct languages.}
\label{sketch}
\end{center}
\end{figure}

Consider now the general case where only $K$ among the $N$ languages survive.
These are necessarily the $K$ most attractive ones,
shown as red symbols in Figure~\ref{sketch}.

In this situation,~(\ref{nfp}) yields
\beq
x_i=\frac{1}{K}+g\,\frac{K+1-2i}{N}
=\frac{1}{K}+g\,\frac{K-2N}{N}+a_i
\label{xim}
\eeq
for $i=1,\dots,K$.
The linear relationship between the attractiveness $a_i$ of language $i$
and the stationary fraction~$x_i$ of speakers of that language,
observed in~(\ref{xin}) and~(\ref{xim}),
is a general feature of the model (see Section~\ref{igal}).
The fraction~$x_K$ of speakers of the least attractive of the surviving languages
vanishes at the following threshold mean attractiveness:
\beq
g_{N,K}=\frac{N}{K(K-1)}
\label{gmdef}
\eeq
for $K=2,\dots,N$.

The following picture therefore emerges for the stationary state of $N$ competing languages
with equally spaced attractivenesses.
The number $K$ of surviving languages decreases as a function of the mean attractiveness $g$,
from $K=N$ (all languages survive) near $g=0$ to $K=1$ (consensus) as very large $g$.
Less attractive languages become extinct one by one
as every single one of the thresholds~(\ref{gmdef}) is traversed, so that
\beq
\matrix{
K=N\hfill &\mbox{for} & 0<g<g_{N,N},\hfill\cr
\dots\hfill\cr
K\;\mbox{(generic)}\hfill\quad &\mbox{for} & g_{N,K+1}<g<g_{N,K},\hfill\cr
\dots\hfill\cr
K=1\hfill &\mbox{for} & g_{N,2}<g<\infty.\hfill\cr
}
\label{msectors}
\eeq

\begin{figure}
\begin{center}
\includegraphics[angle=0,width=.8\linewidth,clip=true]{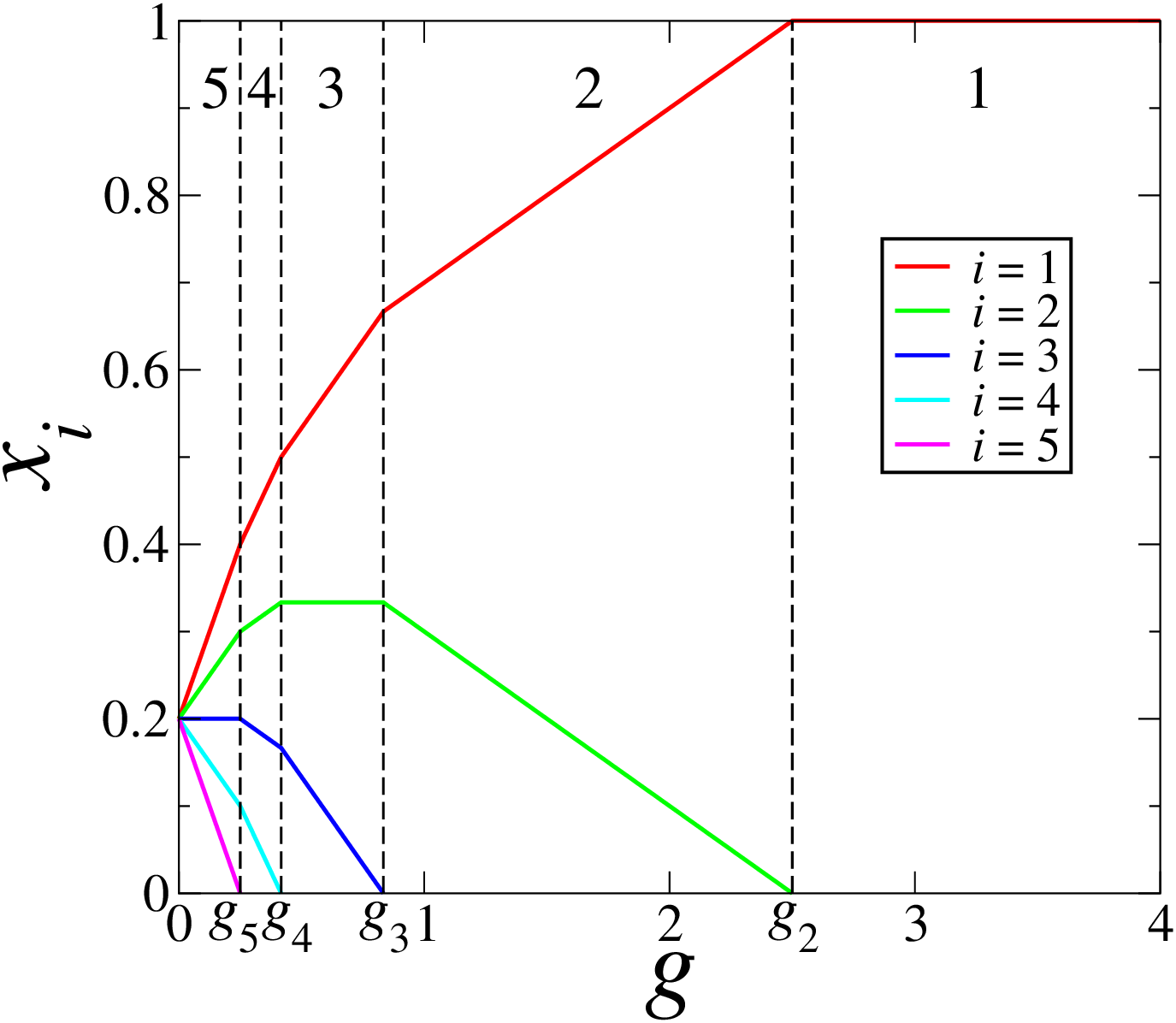}
\caption{
Steady state for 5 competing languages with equally spaced attractivenesses.
The fractions~$x_i$ of speakers of surviving languages
are plotted against the mean attractiveness~$g$ in each sector labelled
by the number $K=1,\dots,5$ of surviving languages.
The threshold values $g_{5,2}=5/2$, $g_{5,3}=5/6$,
$g_{5,4}=5/12$ and $g_{5,5}=1/4$
are abbreviated as $g_2$ to $g_5$.}
\label{plot5}
\end{center}
\end{figure}

Figure~\ref{plot5} illustrates this picture for 5 competing languages.
In each of the sectors defined in~(\ref{msectors}),
the stationary fractions $x_i$ of speakers of the surviving languages are given
by~(\ref{xim}).
They depend continuously on the mean attractiveness $g$,
even though they are given by different expressions in different sectors.
In particular, $x_i$ is flat, i.e., independent of $g$, in the sector where $K=2i-1$.
The fraction $x_1$ of speakers of the most attractive language
grows monotonically as a function of $g$,
whereas all the other fractions of speakers eventually go to zero.

When the number of languages $N$ is large,
the range of values of $g$ where the successive transitions take place is very
broad.
The threshold at which a consensus is reached, $g_{N,2}=N/2$,
is indeed much larger than the threshold at which the least attractive language
disappears,
$g_{N,N}=1/(N-1)$.
The ratio between these two extreme thresholds reads $N(N-1)/2$.

\subsection{The general case}
\label{igal}

We now turn to the general case of $N$ competing languages with arbitrary
reduced attractivenesses~$a_i$.
Throughout the following, languages are numbered in order of decreasing
attractivenesses,~i.e.,
\beq
a_1\ge a_2\ge \dots\ge a_N\ge 0.
\label{aorder}
\eeq

We shall be interested mostly in the stationary state of the model.
As already mentioned above,
the number $K$ of surviving languages
depends on model parameters in a non-trivial way.
The $K$ surviving languages are always the most attractive ones
(see Figure~\ref{sketch}).
The fractions $x_i$ of speakers of those languages,
obeying the fixed-point equations~(\ref{eqfp}),
can be written in full generality as
\beq
x_i=\frac{1-S}{K}+a_i
\label{gxim}
\eeq
for $i=1,\dots,K$, with
\beq
S=\sum_{i=1}^K a_i.
\label{s1}
\eeq
The existence of an explicit expression~(\ref{gxim})
for the solution of the fixed-point equations~(\ref{eqfp}) in full generality
is a consequence of their simple linear-minus-bilinear form,
which also ensures the uniqueness of the attractor.

The number $K$ of surviving languages is the largest such that the
solution~(\ref{gxim}) obeys $x_i>0$
for $i=1,\dots,K$.
Equivalently, $K$ is the largest integer in $1,\dots,N$ such that
\beq
\sum_{i=1}^{K-1}(a_i-a_K)<1.
\eeq
Every single one of the differences involved in the sum is positive, so that:
\beq
\matrix{
K=1:\hfill & a_1-a_2>1,\hfill\cr
K=2:\hfill & a_1+a_2-2a_3>1>a_1-a_2,\hfill\cr
K=3:\hfill & a_1+a_2+a_3-3a_4>1>a_1+a_2-2a_3,\hfill\cr
\dots\hfill\cr
K=N:\hfill & 1>a_1+a_2+\cdots+a_{N-1}-(N-1)a_N.\hfill
}
\label{sectors}
\eeq

From now on, we model attractivenesses as independent random variables.
More precisely, we set
\beq
a_i=w\xi_i,
\eeq
where $w$ is the mean attractiveness,
and the rescaled attractivenesses $\xi_i$
are positive random variables drawn from some continuous distribution $f(\xi)$
such that $\mean{\xi}=1$.
For any given instance of the model,
i.e., any draw of the~$N$ random variables $\{\xi_i\}$,
languages are renumbered in order of decreasing attractivenesses
(see~(\ref{aorder})).

For concreteness we assume that $f(0)$ is non-vanishing
and that $f(\xi)$ falls off more rapidly than $1/\xi^3$ at large $\xi$.
These hypotheses respectively imply that small values of $\xi$ are allowed with
non-negligible probability
and ensure the convergence of the second moment
$\mean{\xi^2}=1+\sigma^2$, where $\sigma^2$ is the variance of $\xi$.

Some quantities of interest can be expressed in closed form for all language
numbers $N$.
One example is the consensus probability $\P$,
defined as the probability of reaching consensus, i.e., of having $K=1$
(see~(\ref{sectors})).
This reads
\beq
\P=\prob{a_1-a_2>1}=\prob{\xi_1-\xi_2>1/w}.
\eeq
We have
\beq
\P=N\int_0^\infty F(\xi)^{N-1} f(\xi+1/w)\,\dd\xi,
\label{pcdef}
\eeq
for all $N\ge2$, where
\beq
F(\xi)=\int_0^\xi f(\xi')\,\dd\xi'
\label{fdef}
\eeq
is the cumulative distribution of $\xi$.

In forthcoming numerical and analytical investigations
we use the following distributions:
\beq
\matrix{
\mbox{Uniform:}\hfill& f(\xi)=\frac{1}{2}\hfill& (0<\xi<2),\hfill\cr
\mbox{Exponential:}\quad& f(\xi)=\e^{-\xi}\hfill& (\xi>0).\hfill\cr
}
\label{dis}
\eeq

\begin{figure}
\begin{center}
\includegraphics[angle=0,width=.8\linewidth,clip=true]{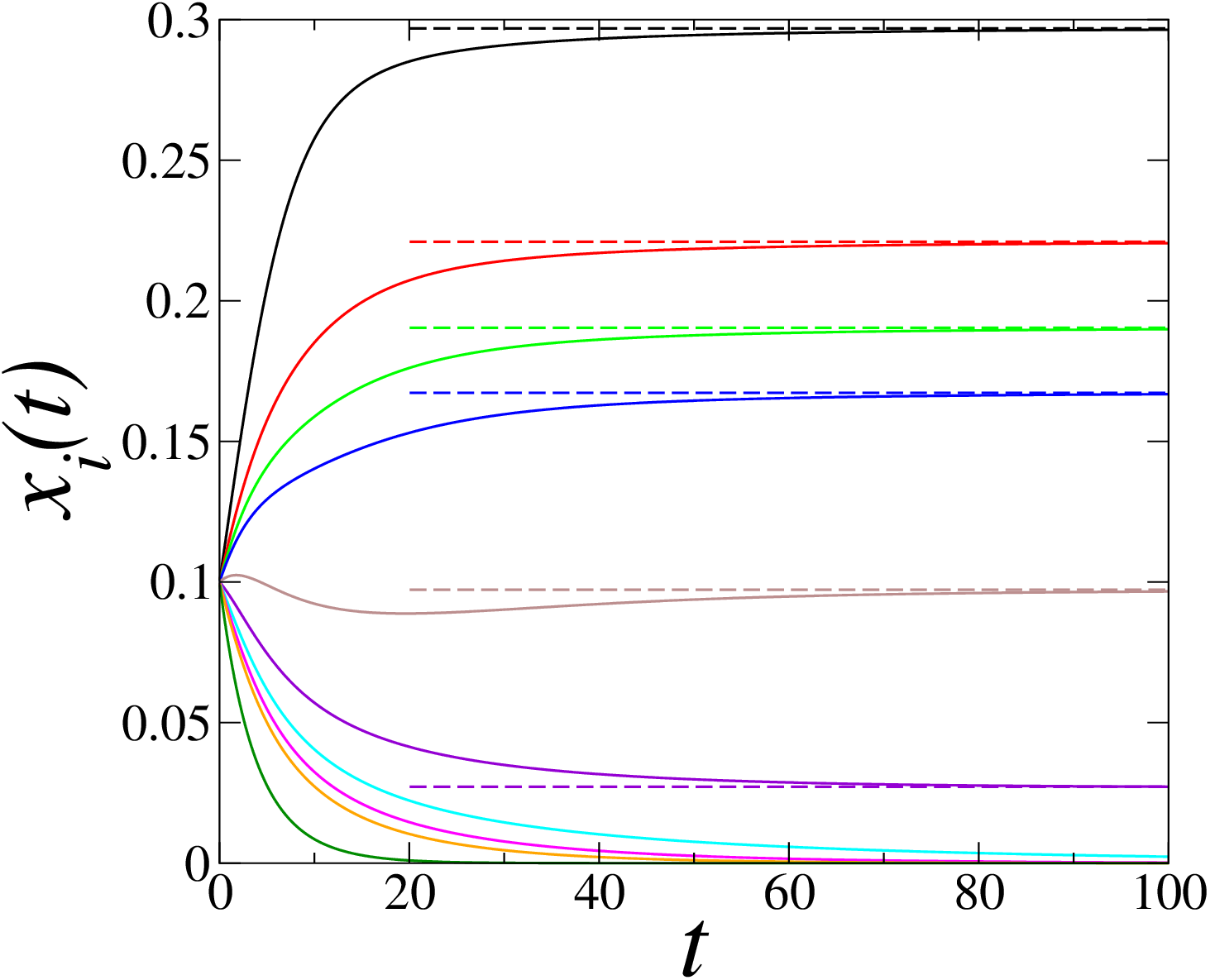}
\caption{
An instance of the model with $N=10$,
a uniform distribution of attractivenesses with $w=0.3$, and $K=6$.
Full curves:
time-dependent fractions of speakers of all languages,
obtained by solving the rate equations~(\ref{eq}) numerically.
Dashed lines: stationary fractions given by~(\ref{gxim}) for $i=1,\dots,6$.}
\label{ux10}
\end{center}
\end{figure}

We begin our exploration of the model
by looking at the dynamics of a typical instance of the model with $N=10$ languages
and a uniform distribution of attractivenesses with $w=0.3$.
Figure~\ref{ux10} shows the time-dependent fractions of speakers of all languages,
obtained by solving the rate equations~(\ref{eq}) numerically,
with the uniform initial condition $x_i(0)=1/10$ for all $i$.
In this example there are $K=6$ surviving languages.
The plotted quantities are observed to converge to their stationary values given
by~(\ref{gxim}) for $i=1,\dots,6$,
and to zero for $i=7,\dots,10$.
They are ordered as the corresponding attractivenesses at all positive times,
i.e., $x_1(t)>x_2(t)>\dots>x_N(t)$.
Some of the fractions however exhibit a non-monotonic evolution.
This is the case for $i=5$ in the present example.

\begin{figure}
\begin{center}
\includegraphics[angle=0,width=.8\linewidth,clip=true]{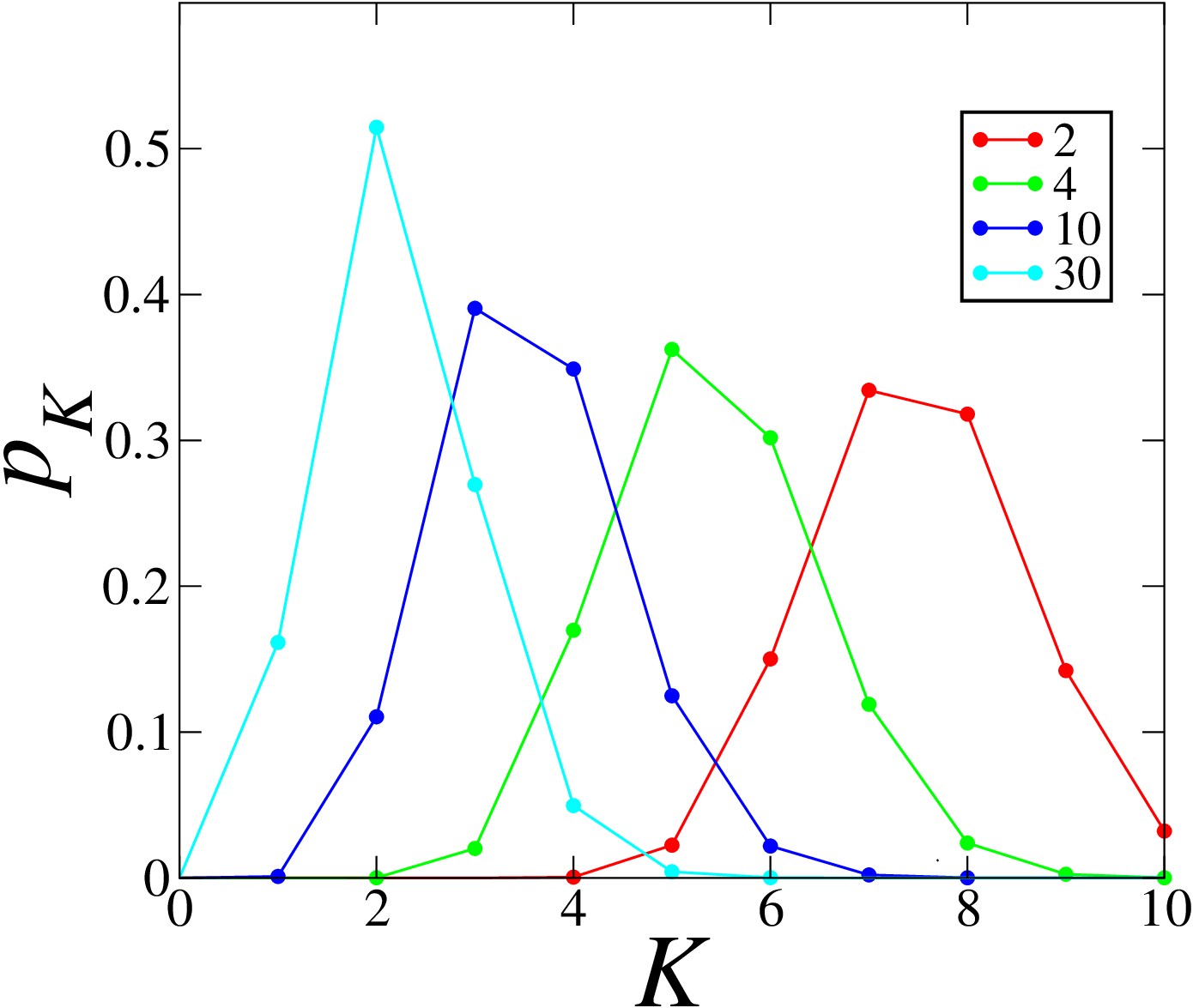}
\vskip 8pt
\includegraphics[angle=0,width=.8\linewidth,clip=true]{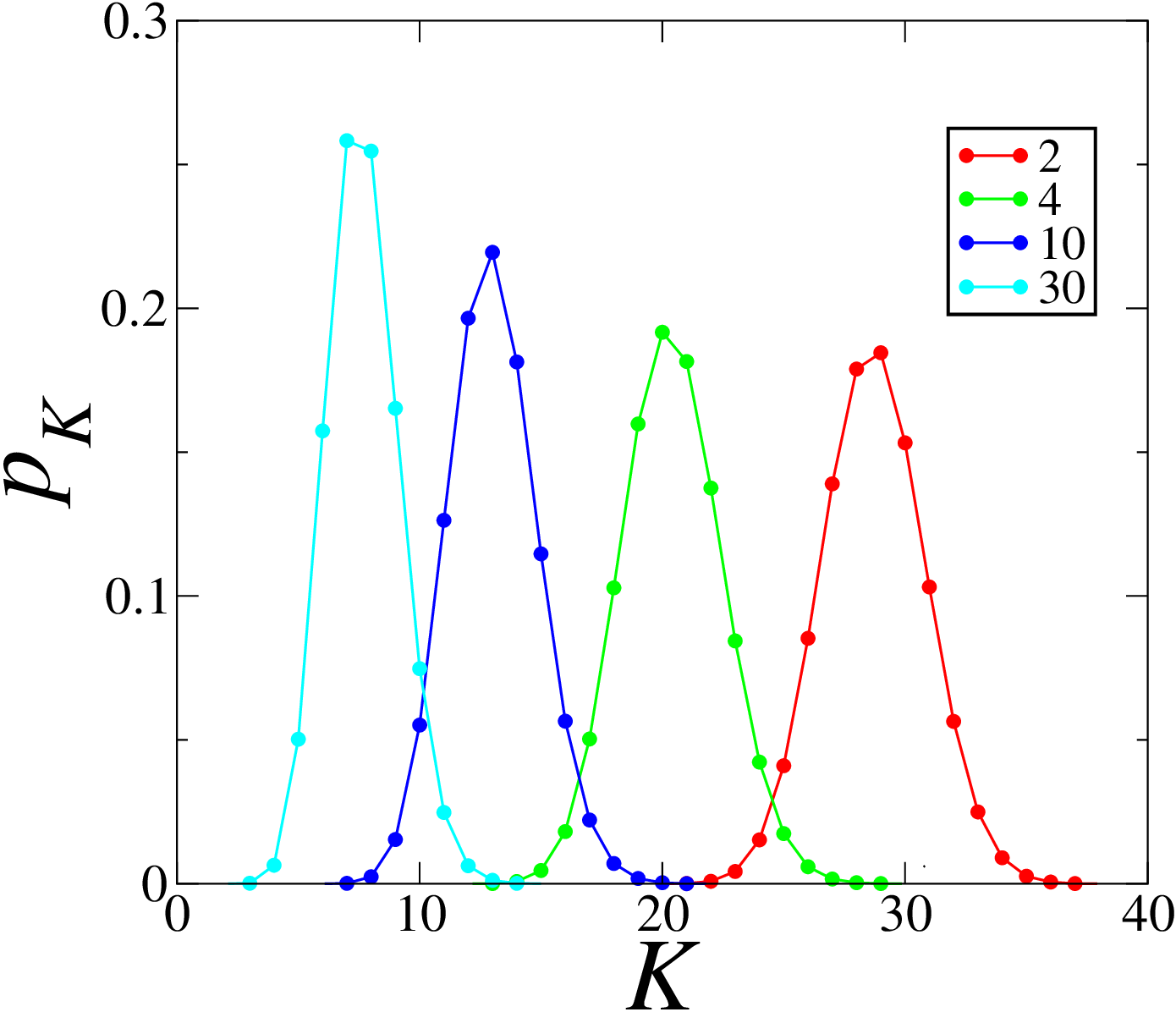}
\caption{
Distribution $p_K$ of the number $K$ of surviving languages,
for $N=10$ (top) and $N=40$ (bottom)
and a uniform distribution of attractivenesses
for four values of $W$ (see legends).}
\label{mh}
\end{center}
\end{figure}

Figure~\ref{mh} shows the distribution $p_K$ of the number $K$ of
surviving languages,
for $N=10$ (top) and $N=40$ (bottom),
and a uniform distribution of attractivenesses for four values of the product
\beq
W=Nw.
\eeq
This choice is motivated by the analysis of Appendix~\ref{appa}.
Each dataset is the outcome of $10^7$ draws of the attractiveness profile.
The widths of the distributions $p_K$ are observed to shrink as $N$ is increased,
in agreement with the expected $1/\sqrt{N}$ behavior stemming from
the law of large numbers.
The corresponding mean fractions $\mean{K}/N$ of surviving languages
are shown in Table~\ref{means} to converge smoothly
to the asymptotic prediction~(\ref{runi}), i.e.,
\beq
\frac{\mean{K}}{N}\to\frac{1}{\sqrt{W}},
\label{scamean}
\eeq
with $1/N$ corrections.

\begin{table}
\begin{center}
\caption
{Mean fraction $\mean{K}/N$ of surviving languages
for a uniform distribution of attractivenesses.
Comparison between numerically measured values for $N=10$ and $N=40$ (see
Figure~\ref{mh})
and the asymptotic analytical prediction~(\ref{scamean}),
for four values of $W$.}
\begin{tabular}{|c|c|c|c|}
\hline
$W$ & $N=10$ & $N=40$ & Eq.~(\ref{scamean})\\
\hline
2 & 0.750 & 0.718 & 0.70711\\
4 & 0.541 & 0.510 & 0.5\\
10 & 0.356 & 0.326 & 0.31623\\
30 & 0.222 & 0.192 & 0.18257\\
\hline
\end{tabular}
\label{means}
\end{center}
\end{table}

\begin{figure}
\begin{center}
\includegraphics[angle=0,width=.8\linewidth,clip=true]{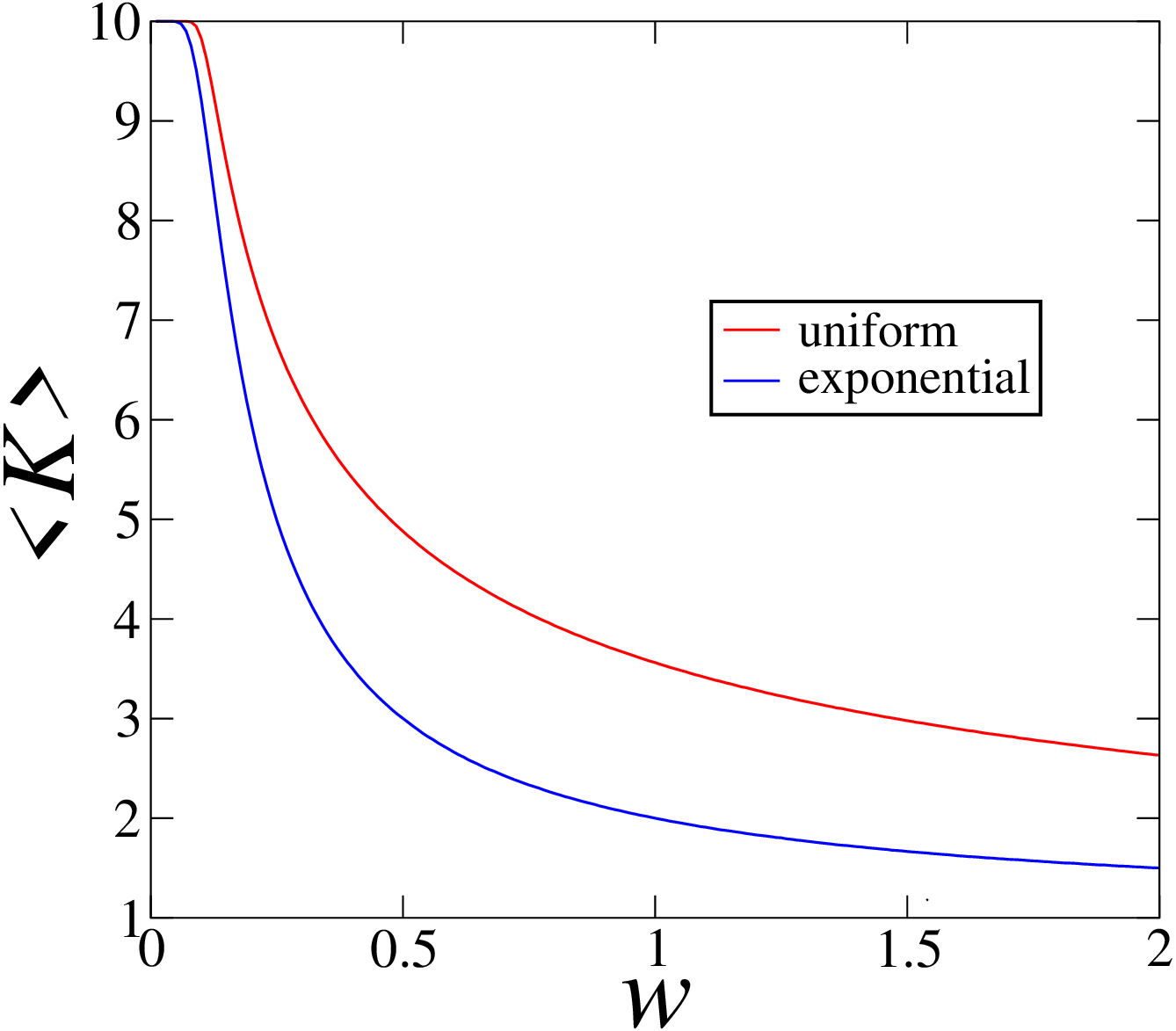}
\caption{
Mean number $\mean{K}$ of surviving languages against mean attractiveness $w$,
for $N=10$ and uniform and exponential attractiveness distributions (see legend).}
\label{ave10}
\end{center}
\end{figure}

An overall picture of the dependence of the statistics of surviving languages
on the mean attractiveness $w$ is provided by Figure~\ref{ave10},
showing the mean number $\mean{K}$ of surviving languages against $w$,
for $N=10$ and uniform and exponential attractiveness distributions.
The plotted quantity decreases monotonically,
starting from the value $\mean{K}=N$ in the absence of conversions ($w=0$),
and converging to its asymptotic value $\mean{K}=1$
in the $w\to\infty$ limit, where consensus is reached with certainty.
Its dependence on $w$ is observed to be steeper for the exponential distribution.
These observations are corroborated by the asymptotic analysis of Appendix~\ref{appa}.
For the uniform distribution,~(\ref{runi}) yields the scaling law
$\mean{K}\approx(N/w)^{1/2}$.
Concomitantly, the consensus probability becomes sizeable for $w\sim N$ (see~(\ref{puni})).
For the exponential distribution,~(\ref{rexp}) yields the decay law
$\mean{K}\approx1/w$, irrespective of $N$,
and the consensus probability is strictly independent of $N$ (see~(\ref{pexp})).

\section{Breaking spatial symmetry:
language coexistence by inhomogeneous attractivenesses}
\label{spatial}

As mentioned in the Introduction,
different competing languages may coexist in distinct geographical areas,
because they are more or less favoured locally,
despite the homogenising effects of migration and language
shift~\cite{pl,ces,fam}.
The aim of this section is to provide a quantitative understanding of this scenario.
We continue to use the approach and the formalism of Section~\ref{internal}.
We however take the liberty of adopting slightly different notations,
as both sections are entirely independent.

\begin{figure}
\begin{center}
\includegraphics[angle=0,width=.7\linewidth,clip=true]{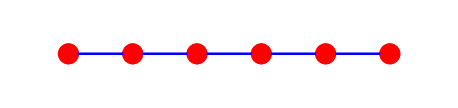}
\caption{
An array of $M=6$ geographical areas.}
\label{array}
\end{center}
\end{figure}

We consider the dynamics of two competing languages in a structured territory
comprising several distinct geographic areas.
For definiteness, we assume that the population of each area is homogeneous.
We restrict ourselves to the geometry of an array of $M$ areas,
where individuals can only migrate along the links joining neighbouring areas,
as shown in Figure~\ref{array}.
We assume for simplicity that the migration rates $\gamma$ between neighbouring areas are uniform,
so that in the very long run single individuals eventually perform
random walks across the territory.
The relative attractivenesses of both competing languages are distributed
inhomogeneously among the various areas,
so that the net conversion rate $C_m$ from language~2 to language~1 depends on
the area number $m$.
Finally, in order to emphasise the effects of spatial inhomogeneity on their own,
we simplify the model by neglecting imbalance and thus set $q=1$.

Let $X_m(t)$ and $Y_m(t)$ denote the respective numbers of speakers of
language~1 and of language~2
in area number $m=1,\dots,M$ at time $t$.
The dynamics of the model is defined by the coupled rate equations
\beqa
\der{X_m(t)}&=&X_m(t)(1-X_m(t)-Y_m(t)+C_mY_m(t))
\nonumber\\
&+&\gamma(X_{m+1}(t)+X_{m-1}(t)-2X_m(t)),
\\
\der{Y_m(t)}&=&Y_m(t)(1-X_m(t)-Y_m(t)-C_mX_m(t))
\nonumber\\
&+&\gamma(Y_{m+1}(t)+Y_{m-1}(t)-2Y_m(t)).
\label{axy}
\eeqa
The extremal sites $m=1$ and $m=M$ have only one neighbour.
The corresponding equations have to be modified accordingly.
The resulting boundary conditions can be advantageously recast as
\beq
X_0(t)=X_1(t),\quad X_{M+1}(t)=X_M(t),
\label{neu}
\eeq
and similarly for other quantities.
These are known as Neumann boundary conditions.

The total populations $P_m(t)=X_m(t)+Y_m(t)$ of the various areas obey
\beqa
\der{P_m(t)}&=&P_m(t)(1-P_m(t))
\nonumber\\
&+&\gamma(P_{m+1}(t)+P_{m-1}(t)-2P_m(t)),
\label{ap}
\eeqa
irrespective of the conversion rates $C_m$.
As a consequence,
in the stationary state
all areas have the same population, which reads $P_m=1$ in our reduced units.
The corresponding stability matrix is given in~(\ref{stap}).
The population profile $P_m(t)$ therefore converges exponentially fast
to its uniform stationary value, with unit relaxation time ($\omega=1$).

From now on we assume, for simplicity, that the total population of each area
is unity in the initial state.
This property is preserved by the dynamics, i.e., we have $P_m(t)=1$ for all
$m$ and $t$,
so that the rate equations~(\ref{axy}) simplify to
\beqa
\der{X_m(t)}&=&C_mX_m(t)(1-X_m(t))
\nonumber\\
&+&\gamma(X_{m+1}(t)+X_{m-1}(t)-2X_m(t)).
\label{ax}
\eeqa

The rate equations~(\ref{ax}) for the fractions $X_m(t)$ of speakers of language~1
in the various areas provide another example
of the broad class of replicator equations
(see e.g.~\cite{hofbook,hofbams,nsscience}).
The above equations are the starting point of the subsequent analysis.
In the situation where language~1 is uniformly favoured or disfavoured,
so that the conversion rates are constant ($C_m=C$),
the above rate equations boil down to the discrete
Fisher-Kolmogo\-rov-Petrovsky-Piscounov
(FKPP) equation~\cite{ZHH,BGL},
which is known to exhibit traveling fronts,
just as the well-known FKPP equation in the continuum~\cite{F,KPP}.
In the present context, the focus will however be
on stationary solutions on finite arrays, obeying
\beq
C_mX_m(1-X_m)+\gamma(X_{m+1}+X_{m-1}-2X_m)=0.
\label{sax}
\eeq

\subsection{Two geographic areas}
\label{stwo}

We begin with the case of two geographic areas connected by a single link.
The problem is simple enough to allow for an explicit exposition of its full
solution.
The rate equations~(\ref{ax}) become
\beqa
\der{X_1(t)}=C_1X_1(t)(1-X_1(t))+\gamma(X_2(t)-X_1(t)),
\\
\der{X_2(t)}=C_2X_2(t)(1-X_2(t))+\gamma(X_1(t)-X_2(t)).
\label{ax2}
\eeqa

Because of the migration fluxes,
for any non-zero $\gamma$ it is impossible for any of the languages
to become extinct in one area and survive in the other one.
The only possibility is that of a uniform consensus,
where one and the same language survives in all areas.
The consensus state where language~1 survives
is described by the stationary solution $X_1=X_2=1$.
The corresponding stability matrix is
\beq
\m S_2^\un=\pmatrix{-C_1-\gamma &\gamma\cr\gamma & -C_2-\gamma}
=-\diag(C_1,C_2)-\gamma\m\Delta_2,
\label{s2un}
\eeq
where $\diag(\dots)$ denotes a diagonal matrix (whose entries are listed),
whereas $\m\Delta_2$ is defined in~(\ref{lap}).
The stability condition amounts to
\beq
C_1+C_2+2\gamma>0,\quad C_1C_2+\gamma(C_1+C_2)>0.
\label{cdun}
\eeq
Similarly, the consensus state where language~2 survives
is described by the stationary solution $X_1=X_2=0$.
The corresponding stability matrix is
\beq
\m S_2^\ze=\pmatrix{C_1-\gamma &\gamma\cr\gamma & C_2-\gamma}
=\diag(C_1,C_2)-\gamma\m\Delta_2.
\label{s2ze}
\eeq
The conditions for the latter to be stable read
\beq
C_1+C_2-2\gamma<0,\quad C_1C_2-\gamma(C_1+C_2)>0.
\label{cdze}
\eeq

\begin{figure}
\begin{center}
\includegraphics[angle=0,width=.7\linewidth,clip=true]{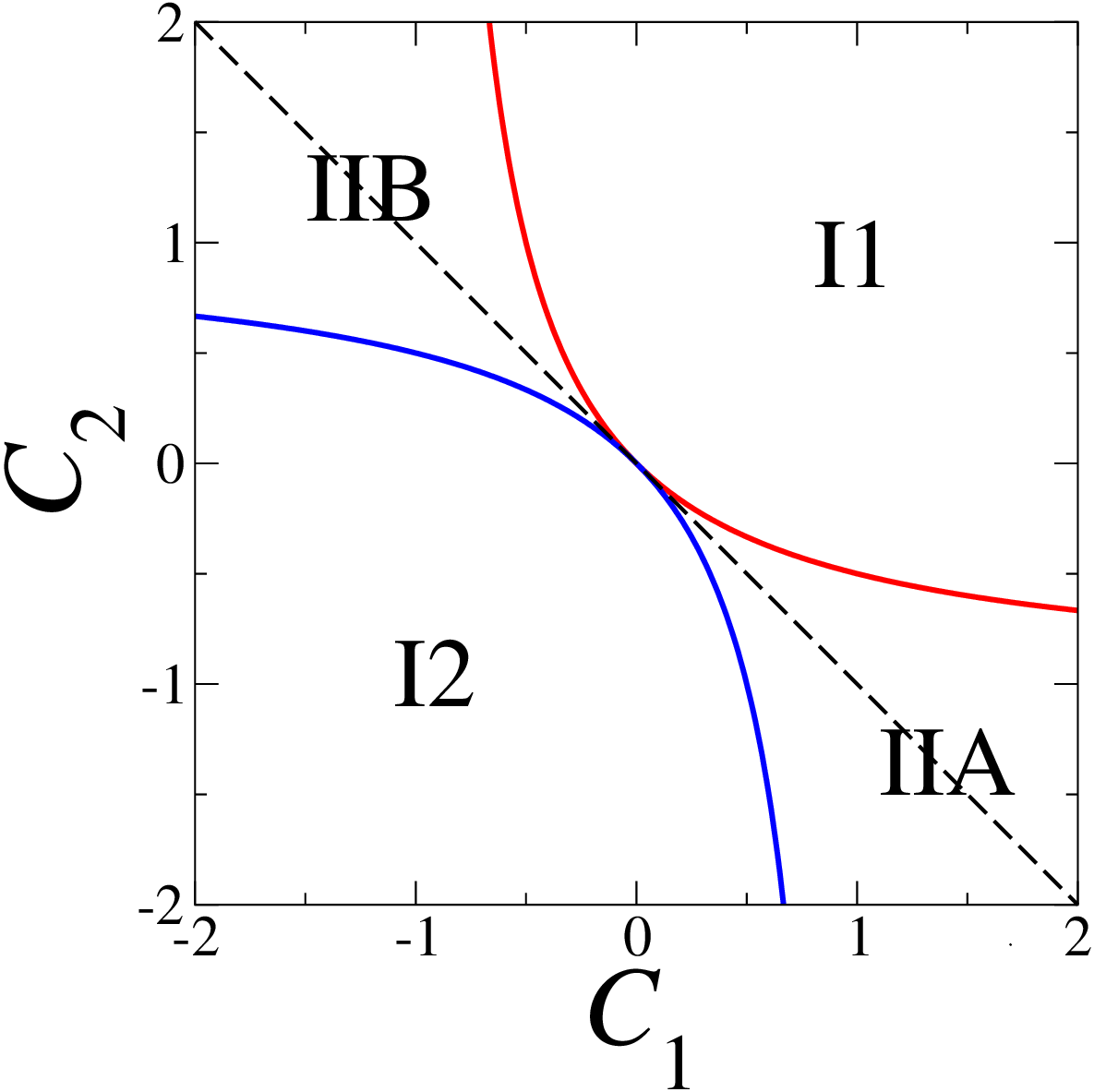}
\caption{
Phase diagram in the $C_1$--$C_2$ plane
of the model defined on two geographic areas for $\gamma=1$.
I1: consensus phase where language~1 survives.
I2: consensus phase where language~2 survives.
IIA and IIB: coexistence of both languages in both areas.
Black dashed line: $C_1+C_2=0$ (none of the languages is globally favoured).}
\label{c1c2plot}
\end{center}
\end{figure}

Figure~\ref{c1c2plot} shows the phase diagram of the model in the $C_1$--$C_2$
plane for $\gamma=1$.
Region I1 is the consensus phase where language~1 survives.
It is larger than the quadrant where this language is everywhere favoured
(i.e., $C_1$ and $C_2$ are positive),
as its boundary (red curve) reads $C_1C_2+\gamma(C_1+C_2)=0$.
Similarly,
region I2 is the consensus phase where language~2 survives.
It is larger than the quadrant where this language is everywhere favoured
(i.e., $C_1$ and $C_2$ are negative),
as its boundary (blue curve) reads $C_1C_2-\gamma(C_1+C_2)=0$.
The regions marked IIA and IIB are coexistence phases.
These phases are located symmetrically around the line $C_1+C_2=0$ (black
dashed line)
where none of the languages is globally favoured.
There, the fractions $X_1$ and $X_2$ of speakers of language~1 in both areas
vary continuously between zero on the blue curve and unity on the red one,
according to
\beqa
\hbox{IIA:}\quad&
X_1=\frad{1}{2}-\frad{\gamma}{C_1}-D,\quad
X_2=\frad{1}{2}-\frad{\gamma}{C_2}+D,
\\
\hbox{IIB:}\quad&
X_1=\frad{1}{2}-\frad{\gamma}{C_1}+D,\quad
X_2=\frad{1}{2}-\frad{\gamma}{C_2}-D,
\eeqa
with
\beq
D=\sqrt{\frac{1}{4}-\frac{\gamma^2}{C_1C_2}}.
\eeq
We have therefore
\beq
X_1+X_2=1-\gamma\,\frac{C_1+C_2}{C_1C_2}
\eeq
all over the coexistence phases IIA and IIB.
The right-hand-side equals~0 on the blue curve, 1 on the black dashed line, and
2 on the red curve.

\subsection{$M$ geographical areas}
\label{sm}

From now on we consider the general situation of $M$ geographic areas,
as shown in Figure~\ref{array}.
The basic properties of the model can be inferred from the case of two areas,
studied in section~\ref{stwo}.
In full generality, because of migration fluxes,
it is impossible for any of the languages to become extinct in some areas
and survive in some other ones.
The only possibility is that of a uniform consensus,
where one and the same language survives in all areas.

The consensus state where language~1 survives
is described by the uniform stationary solution where $X_m=1$ for all $m=1,\dots,M$.
The corresponding stability matrix~is
\beq
\m S_M^\un=-\diag(C_1,\dots,C_M)-\gamma\m\Delta_M.
\label{sun}
\eeq
Similarly, the consensus state where language~2 survives
corresponds to the stationary solution where $X_m=0$ for all $m=1,\dots,M$.
The corresponding stability matrix is
\beq
\m S_M^\ze=\diag(C_1,\dots,C_M)-\gamma\m\Delta_M.
\label{sze}
\eeq
These expressions respectively generalise~(\ref{s2un}) and~(\ref{s2ze}).

If all the conversion rates $C_m$ vanish,
both the above matrices read $-\gamma\m\Delta_M$,
whose spectrum comprises one vanishing eigenvalue (see~(\ref{deltaspec})).
In the regime where all the conversion rates $C_m$ are small with respect to $\gamma$,
perturbation theory tells us that the largest eigenvalues
of $\m S_M^\ze$ and $\m S_M^\un$ respectively read $\overline C$ and $-\overline C$,
to leading order,
where
\beq
\overline C
=\m\phi_0\cdot\diag(C_1,\dots,C_M)\m\phi_0
=\frac{1}{M}\sum_{m=1}^M C_m.
\label{barc}
\eeq
We therefore predict that the average conversion rate $\overline C$
determines the fate of the system in the regime where conversion rates are
small with respect to $\gamma$.
If language~1 is globally favoured, i.e., $\overline C>0$,
the system reaches the consensus where language~1 survives, and vice versa.

In the generic situation where the conversion rates $C_m$ are comparable to~$\gamma$,
their dispersion around their spatial average $\overline C$
broadens the spectra of the matrices $\m S_M^\un$ and $\m S_M^\ze$.
As a consequence, the condition $\overline C>0$ (resp.~$\overline C<0$) is necessary, albeit not
sufficient, for the consensus where language~1 (resp.~language~2) survives to be stable.

In the following we shall successively consider
ordered attractiveness profiles in Section~\ref{sorder}
and random ones in Section~\ref{srandom}.

\subsection{Ordered attractiveness profiles}
\label{sorder}

\begin{figure}
\begin{center}
\includegraphics[angle=0,width=.7\linewidth,clip=true]{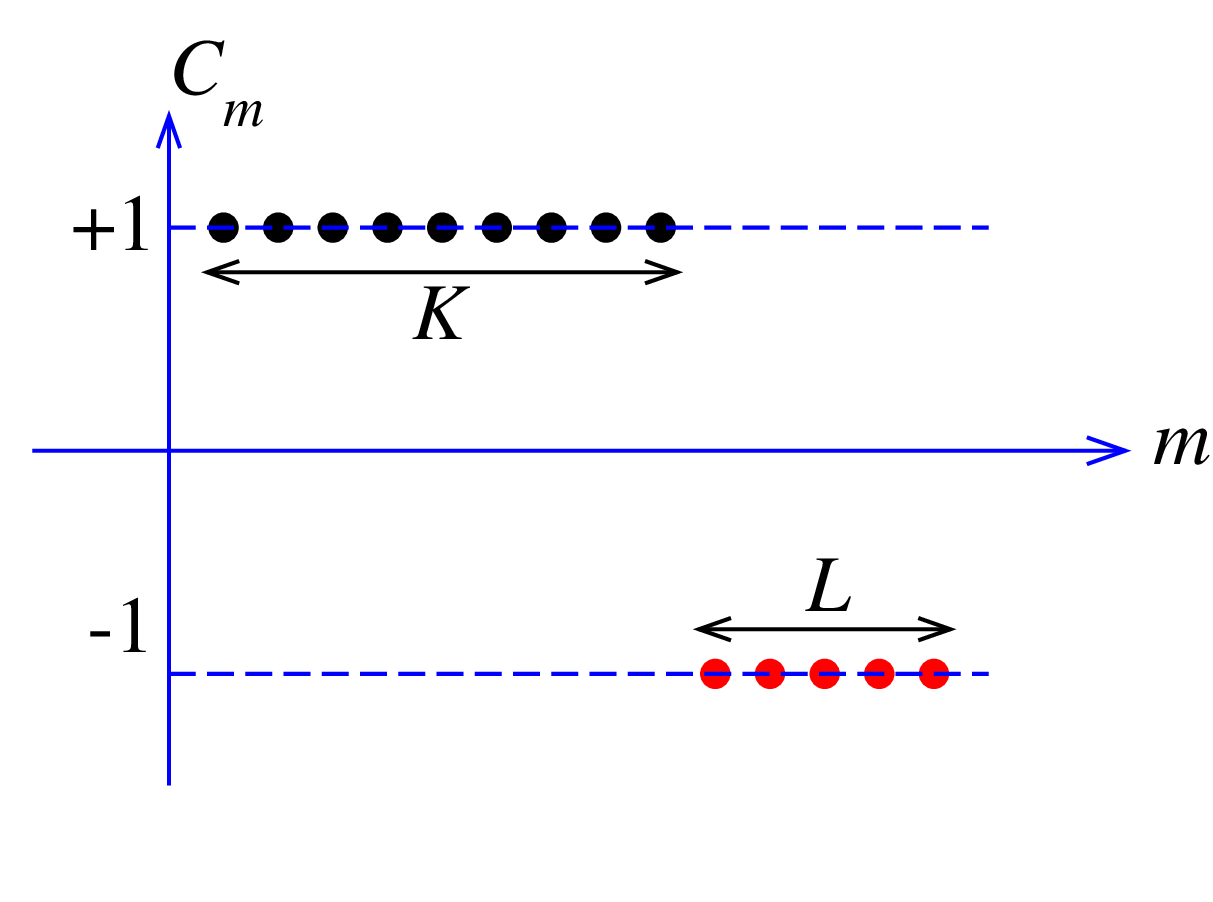}
\caption{
The ordered profile of conversion rates defined in~(\ref{cord}).}
\label{order}
\end{center}
\end{figure}

This section is devoted to a simple situation where the attractiveness profiles
of both languages are ordered spatially.
More specifically, we consider the case
where language~1 is favoured in the $K$ first (i.e., leftmost) areas,
whereas language~2 is favoured in the $L$ last (i.e., rightmost) areas,
with $K\ge L$ and $K+L=M$.
For the sake of simplicity,
we choose to describe this situation by conversion rates that have unit magnitude,
as shown in Figure~\ref{order}:
\beq
C_m=\left\{\matrix{
+1\hfill&\hbox{for}& m=1,\dots,K,\hfill\cr
-1\hfill&\hbox{for}& m=K+1,\dots,M.
}\right.
\label{cord}
\eeq

\begin{figure}
\begin{center}
\includegraphics[angle=0,width=.8\linewidth,clip=true]{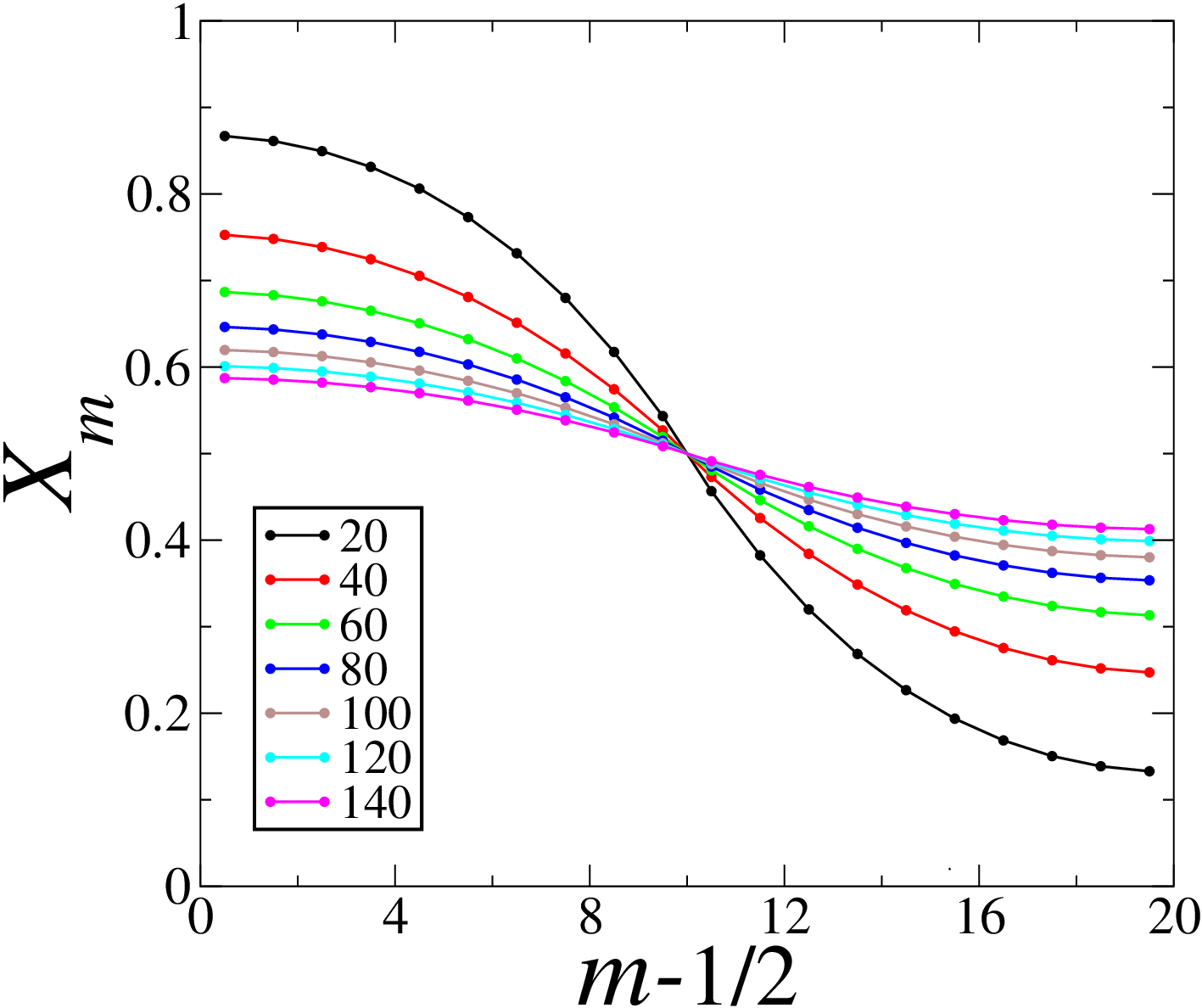}
\vskip 6pt
\includegraphics[angle=0,width=.8\linewidth,clip=true]{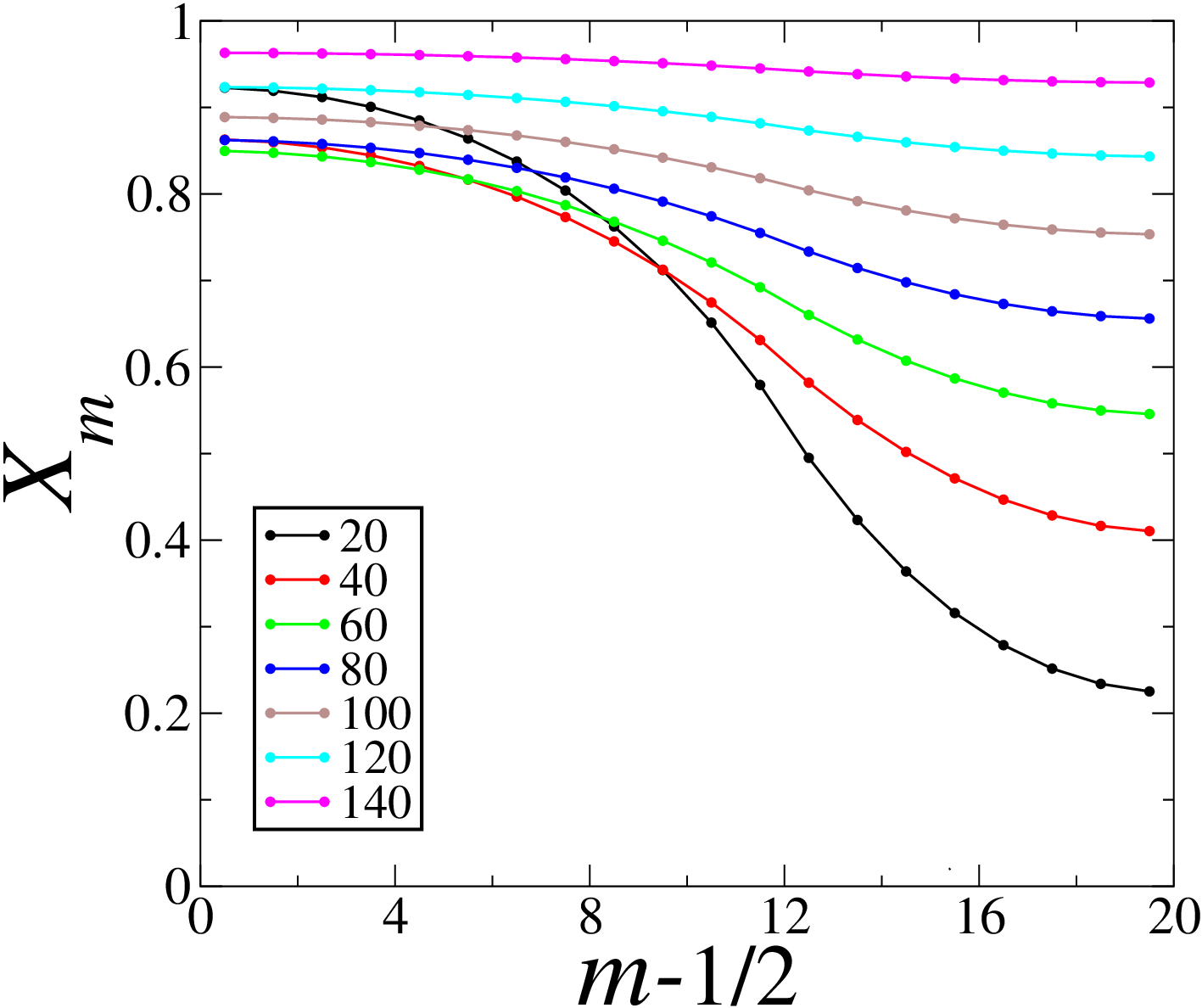}
\caption{
Stationary fraction $X_m$ of speakers of language~1 against $m-1/2$
in two cases of ordered attractiveness profiles on an array of $M=20$ areas,
for several migration rates $\gamma$ (see legends).
Top: symmetric situation where $K=L=10$.
Bottom: asymmetric situation where $K=12$ and $L=8$.}
\label{xor}
\end{center}
\end{figure}

The symmetric situation where $M$ is even and $K=L=M/2$,
so that $\overline C=0$, can be viewed as a generalisation
of the case of two geographic areas, studied in Section~\ref{stwo},
for $C_1+C_2=0$, i.e., along the black dashed line of Figure~\ref{c1c2plot}.
Both languages play symmetric roles,
so that no language is globally preferred, and no consensus can be reached.
As a consequence, both languages survive everywhere,
albeit with non-trivial spatial profiles,
which can be thought of as avatars of the FKPP traveling fronts mentioned above,
rendered stationary by being pinned by boundary conditions.
The upper panel of Figure~\ref{xor} shows the stationary fraction $X_m$
of speakers of language~1
against area number, for $M=20$ (i.e., $K=L=10$) and several $\gamma$.
The abscissa $m-1/2$ is chosen in order to have a symmetric plot.
As one might expect, each language is preferred in the areas where it is favoured,
i.e., we have $X_m>1/2$ for $m=1,\dots,K$,
whereas $X_m<1/2$ for $m=K+1,\dots,M$.
Profiles get smoother as the migration rate $\gamma$ is increased.
The width~$\xi$ of the transition region
is indeed expected to grow as
\beq
\xi\sim\sqrt\gamma.
\eeq
This scaling law is nothing but the large $\gamma$ behaviour
of the exact dispersion relation
\beq
4\gamma\sinh^2\muh=1
\eeq
(see~(\ref{imagdisp}))
between $\gamma$ and the decay rate $\mu$
such that either $X_m$ or $1-X_m$ falls off as $\e^{\pm m\mu}$,
with the natural identification $\xi=1/\mu$.

The asymmetric situation where $K>L$,
so that $\overline C=(K-L)/M>0$,
implying that language~1 is globally favoured, is entirely different.
The system indeed reaches a consensus state
where the favoured language survives,
whenever the migration rate $\gamma$ exceeds some threshold $\gamma_c$.
This threshold, corresponding to the consensus state becoming marginally stable,
only depends on the integers~$K$ and $L$.
It is derived in Appendix~\ref{appb} and given by the largest solution of~(\ref{tata}).

This is illustrated in the lower panel of Figure~\ref{xor},
showing $X_m$ against $m-1/2$ for $K=12$ and $L=8$,
and the same values of $\gamma$ as on the upper panel.
The corresponding threshold reads $\gamma_c=157.265$.
The whole profile shifts upwards while it broadens as $\gamma$ is increased.
It tends uniformly to unity as $\gamma$ tends to $\gamma_c$,
demonstrating the continuous nature of the transition where consensus is formed.

The threshold migration rate $\gamma_c$ assumes a scaling form
in the regime where $K$ and $L$ are large and comparable.
Setting
\beq
K=\frac{1+f}{2}\,M,\quad L=\frac{1-f}{2}\,M,
\eeq
so that the excess fraction $f$ identifies with the average conversion rate
$\overline C$,
the threshold rate $\gamma_c$ grows quadratically with the system size $M$,
according to
\beq
\gamma_c\approx\frac{M^2}{4g(f)^2},
\label{gammaorder}
\eeq
where $g(f)$ is the smallest positive solution of the implicit equation
\beq
\tanh((1+f)g(f))=\tan((1-f)g(f)),
\eeq
which is a rescaled form of~(\ref{tata}).

The quadratic growth law~(\ref{gammaorder})
is a consequence of the diffusive nature of migrations.
The following limiting cases deserve special mention.

\noindent For $f\to0$, i.e., $K$ and $L$ relatively close to each other
($K-L\ll M$),
we have
\beq
g(f)=\sqrt{3f}\left(1+\frac{27}{35}\,f^2+\cdots\right),
\eeq
yielding to leading order
\beq
\gamma_c\approx\frac{M^3}{12(K-L)}.
\eeq

\noindent For $f\to1$, i.e., $L\ll K$, we have $g(f)\approx\pi/(4(1-f))$,
up to exponentially small corrections, so that
\beq
\gamma_c\approx\frac{16L^2}{\pi^2}.
\eeq

The situation considered in the lower panel of Figure~\ref{xor},
i.e., $M=20$, $K=12$ and $L=8$, corresponds to $f=1/5$,
hence $g=0.799622814\dots$, so that
\beq
\gamma_c\approx 0.390993606\dots M^2.
\eeq
This scaling result predicts $\gamma_c\approx156.397$ for $M=20$,
a good approximation to the exact value $\gamma_c=157.265$.

\subsection{Random attractiveness profiles}
\label{srandom}

We now consider the situation of randomly disordered attractiveness profiles.
The conversion rates $C_m$ are modelled as independent random variables
drawn from some symmetric distribution $f(C)$,
such that $\mean{C_m}=0$ and $\mean{C_m^2}=w^2$.

The first quantity we will focus on is the consensus probability $\P$.
It is clear from a dimensional analysis of the rate equations~(\ref{sax})
that $\P$ depends on the ratio $\gamma/w$, the system size $M$,
and the distribution $f(C)$.
Furthermore,~$\P$ is expected to increase with $\gamma/w$.
It can be estimated as follows in the limiting situations where $\gamma/w$
is either very small or very large.

In the regime where $\gamma\ll w$
(e.g.~far from the center in Figure~\ref{c1c2plot}),
conversion effects dominate migration effects.
There, a consensus where language~1 (resp.~language~2) survives can only be reached
if all conversion rates $C_m$ are positive (resp.~negative).
The total consensus probability thus scales as
\beq
\P\approx\frac{1}{2^{M-1}}.
\eeq
Consensus is therefore highly improbable in this regime.
In other words, coexistence of both languages
is overwhelmingly the rule.

In the opposite regime where $\gamma\gg w$
(e.g.~in the vicinity of the center in Figure~\ref{c1c2plot}),
migration effects dominate conversion effects.
There, we have seen in Section~\ref{sm}
that the average conversion rate defined in~(\ref{barc})
essentially determines the fate of the system.
If language~1 is globally favoured, i.e., $\overline C>0$,
then the system reaches the uniform consensus where language~1 survives, and vice versa.
Coexistence is therefore rare in this regime,
as it requires~$\overline C$ to be atypically small.
The probability $\Q$ for this to occur, to be identified with $1-\P$,
has been given a precise definition in Appendix~\ref{appb}
by means of the expansion~(\ref{dsmall}) of $D_M=\det\m S_M^\un$ as a power series in the $C_m$,
and estimated within a simplified Gaussian setting.
In spite of the heuristic character of its derivation,
the resulting estimate~(\ref{qrough}) demonstrates that the consensus
probability scales as
\beq
\P\approx\Phi(x),\quad x=\frac{\gamma}{M^{3/2}w}
\label{psca}
\eeq
all over the regime where the ratio $\gamma/w$ and the system size~$M$ are both large.
Furthermore, taking~(\ref{qrough}) literally,
we obtain the following heuristic prediction for the finite-size scaling function:
\beq
\Phi_{\rm heuristic}(x)=\frac{2}{\pi}\arctan(x\sqrt{12}).
\label{fheu}
\eeq

The scaling result~(\ref{psca}) shows that the scale of the migration rate $\gamma$
which is relevant to describe the consensus probability for a typical disordered
profile of attractivenesses reads
\beq
\gamma\sim M^{3/2}w.
\label{gammadis}
\eeq
This estimate grows less rapidly with $M$ than the corresponding threshold for ordered profiles,
which obeys a quadratic growth law (see~(\ref{gammaorder})).
The exponent $3/2$ of the scaling law~(\ref{gammadis})
can be put in perspective with the anomalous scaling of the localisation length
in one-dimensional Anderson localisation near band edges.
There is indeed a formal analogy between the stability matrices of the present problem
and the Hamiltonian of a tight-binding electron in a disordered potential,
with the random conversion rates $C_m$ replacing the disordered on-site energies.
For the tight-binding problem,
the localisation length is known to diverge as $\xi\sim1/w^2$ in the bulk of the spectrum,
albeit only as $\xi\sim1/w^{2/3}$ in the vicinity of band
edges~\cite{flo,hal,dgedge,irt,cltt}.
Replacing $\xi$ by the system size $M$ and remembering that~$w$ stands for~$w/\gamma$,
we recover~(\ref{gammadis}).
The exponent $3/2$ is therefore nothing but the inverse of the exponent
$2/3$ of anomalous band-edge localisation.

\begin{figure}
\begin{center}
\includegraphics[angle=0,width=.8\linewidth,clip=true]{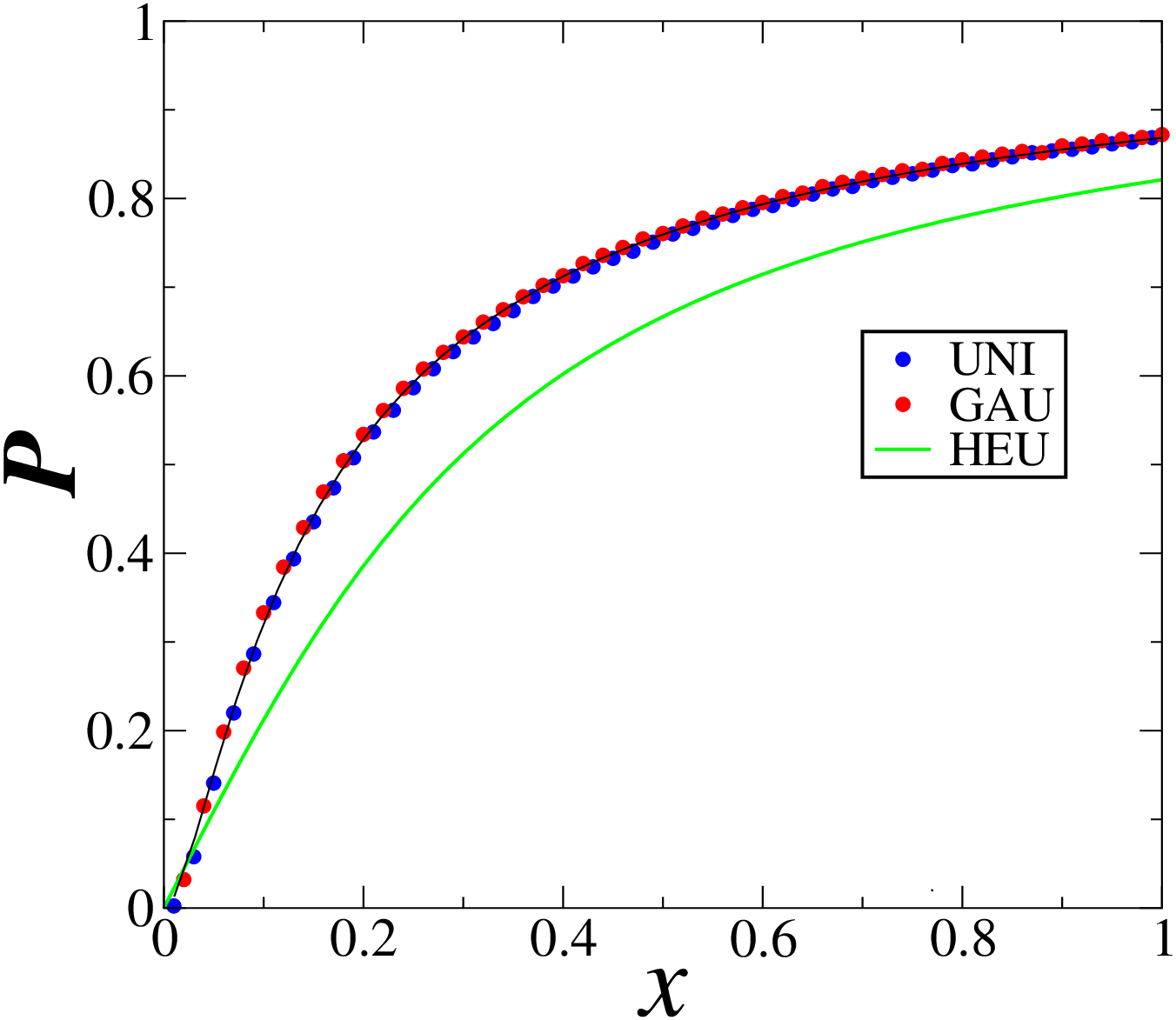}
\caption{
Finite-size scaling plot of the consensus probability $\P$ against $x=\gamma/M^{3/2}$.
Symbols: data for $M=20$ and uniform (UNI) and Gaussian (GAU)
conversion rate distributions with $w=1$.
Thin black curve: guide to the eye pointing toward the universality
of the finite-size scaling function $\Phi$ entering~(\ref{psca}).
Full green curve: heuristic (HEU) prediction~(\ref{fheu}).}
\label{fss}
\end{center}
\end{figure}

Figure~\ref{fss} shows a finite-size scaling plot
of the consensus probability $\P$ against $x=\gamma/M^{3/2}$.
Data correspond to arrays of length $M=20$
with uniform and Gaussian distributions of conversion rates with $w=1$.
Each data point is the outcome of $10^6$ independent realisations.
The thin black curve is a guide to the eye,
suggesting that the finite-size scaling function $\Phi$ is universal,
i.e., independent of details of the conversion rate distribution.
It has indeed been checked that the weak residual dependence of data points
on the latter distribution
becomes even smaller as~$M$ is further increased.
The full green curve shows the heuristic prediction~(\ref{fheu}),
providing a semi-quantitative picture of the finite-size scaling function.
For instance, consensus is reached with probability $\P=1/2$ and $\P=2/3$
respectively for $x\approx0.18$ and $x\approx0.33$,
according to actual data,
whereas~(\ref{fheu}) respectively predicts $x=1/\sqrt{12}=0.288675\dots$ and $x=1/2$.

\begin{figure}
\begin{center}
\includegraphics[angle=0,width=.8\linewidth,clip=true]{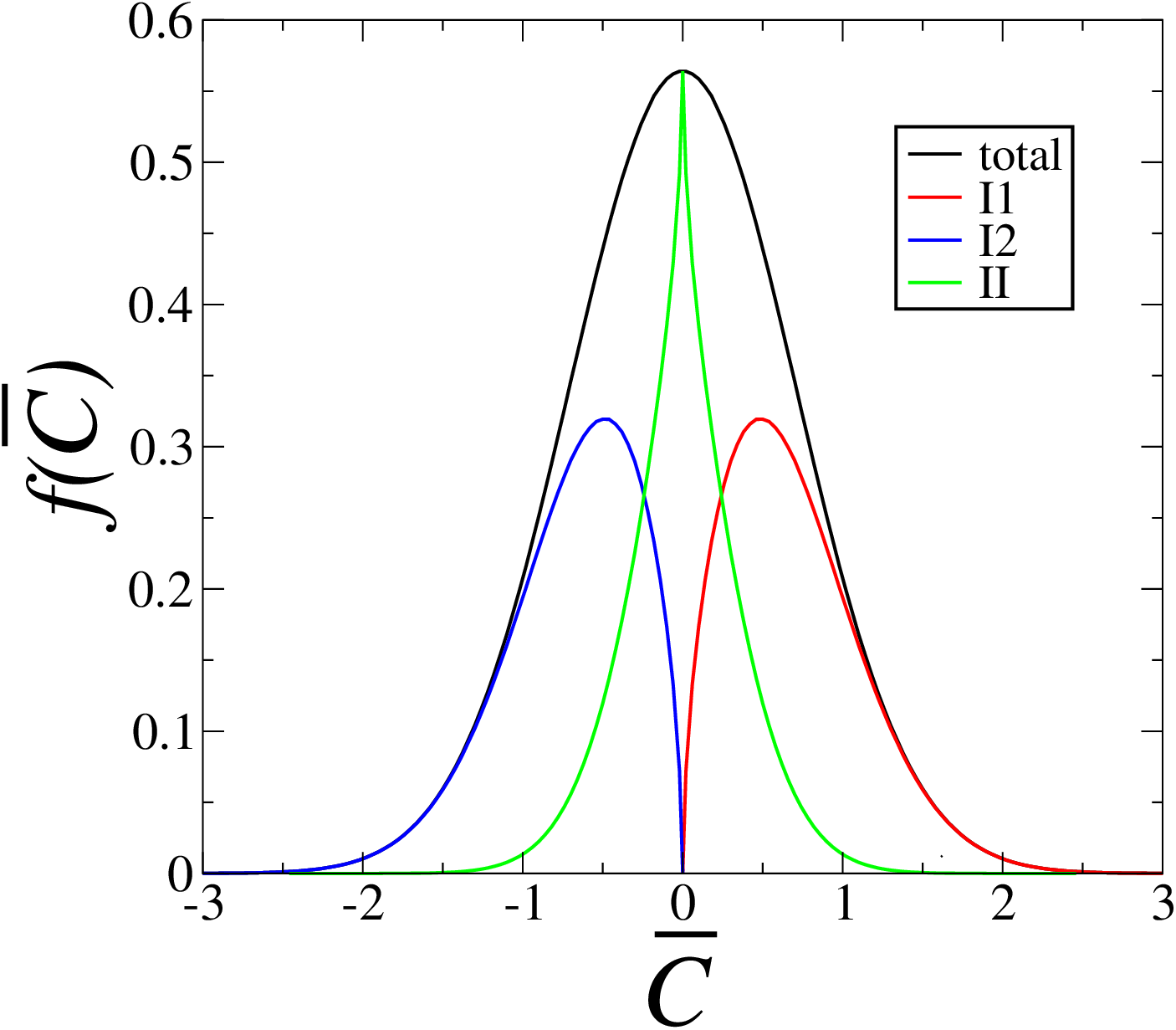}
\vskip 6pt
\includegraphics[angle=0,width=.8\linewidth,clip=true]{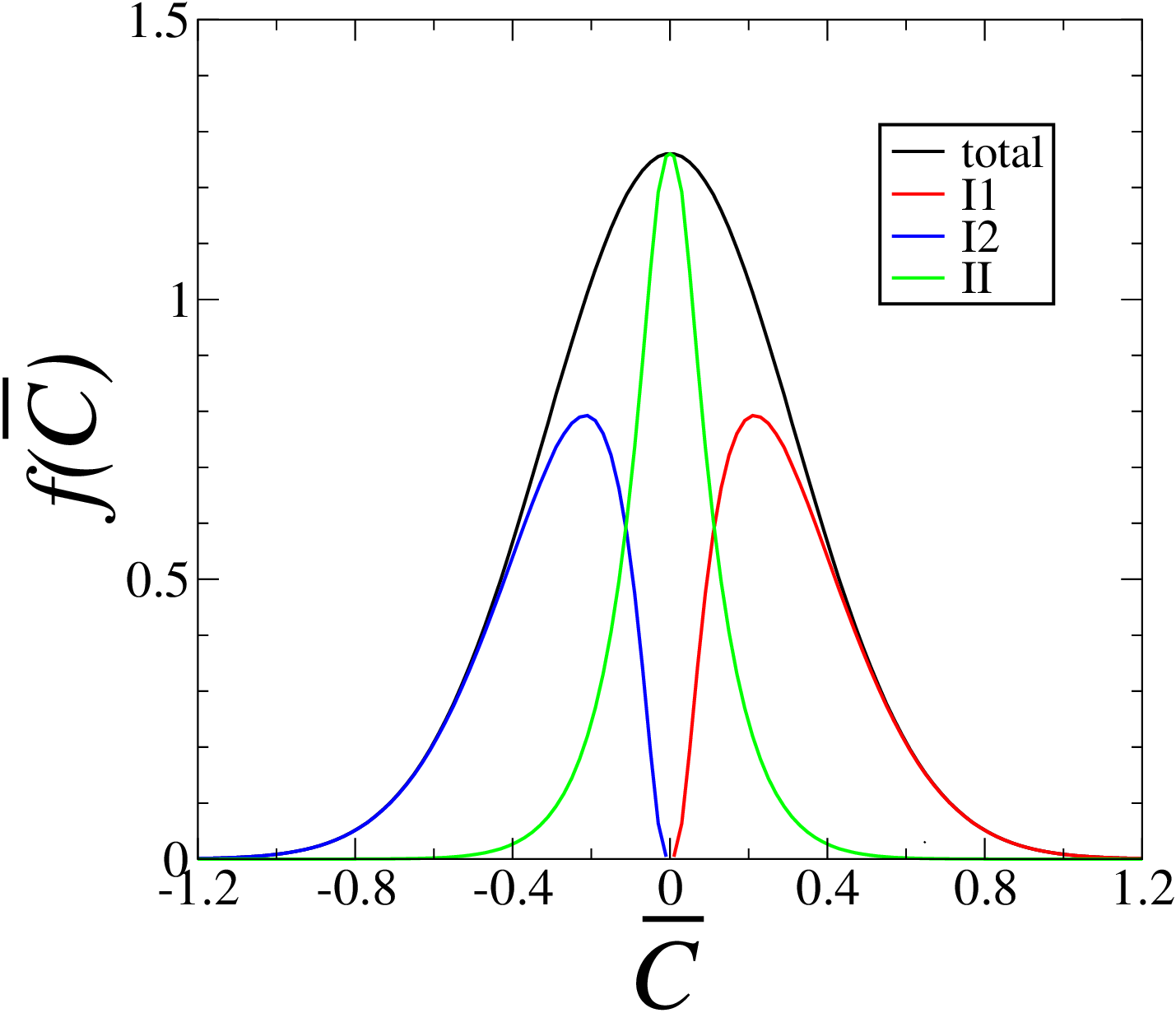}
\caption{
Probability distribution of the average conversion rate $\overline C$
for a Gaussian distribution of conversion rates with $w=1$.
Black curves: total (i.e., unconditioned) distribution.
Red curves: distribution conditioned on consensus I1.
Blue curves: distribution conditioned on consensus I2.
Green curves: distribution conditioned on coexistence (II).
Top: $M=2$ and $\gamma=0.351$.
Bottom: $M=10$ and $\gamma=10.22$.}
\label{h}
\end{center}
\end{figure}

Besides the value of the consensus probability $\P$,
the next question is what determines whether or not the system reaches consensus.
In Section~\ref{sm}
it has been demonstrated that the average conversion rate $\overline C$
defined in~(\ref{barc}) essentially determines the fate of the system
in the regime where migration effects dominate conversion effects.
It has also been shown
that the consensus denoted by I1, where language~1 survives,
can only be stable for $\overline C>0$,
whereas the consensus denoted by I2, where language~2 survives,
can only be stable for $\overline C<0$.
The above statements are made quantitative in Figure~\ref{h},
showing the probability distribution
of the average conversion rate $\overline C$,
for a Gaussian distribution of conversion rates with $w=1$.
The total (i.e., unconditioned) distribution (black curves) is Gaussian.
Red and blue curves show the distributions conditioned on consensus.
They are indeed observed to live entirely on $\overline C>0$ for I1
and on $\overline C<0$ for I2.
Finally, the distributions conditioned on coexistence (green curves, denoted by II)
exhibit narrow symmetric shapes around the origin.
Values of the migration rate $\gamma$ are chosen so as to have three partial histograms
with equal weights, i.e., a consensus probability $\P=2/3$.
This fixes $\gamma\approx0.351$ for $M=2$ (top)
and $\gamma\approx10.22$ for $M=10$ (bottom).

\section{Discussion}
\label{disc}

An area of interest that is common to both physicists and linguists
concerns the evolution of competing languages.
It was long assumed that such competition would result in the dominance
of one language above all its competitors,
until some recent work hinted that coexistence might be possible
under specific circumstances.
We argue here that coexistence of two or more competing languages can result from two symmetry-breaking mechanisms
-- due respectively to {\it imbalanced internal dynamics}
and {\it spatial heterogeneity} -- and engage in a
quantitative exploration of the circumstances
which lead to this coexistence.
In this work,
both symmetry-breaking scenarios are dealt with on an equal footing.

In the first case of competing languages in a single geographical area,
our introduction of an interpolation parameter $q$, which
measures the amount of imbalance in the internal dynamics,
turns out to be crucial for the investigation of language coexistence.
It is conceptually somewhat subtle, since it appears only in the saturation terms in the coupled logistic equations
used here to describe language competition; in contrast to the conversion terms (describing
language shift
from a less to a more favoured language), its appearance is {\it symmetric} with respect to both languages.
For multiply many competing languages,
the ensuing rate equations for the fractions of speakers
are seen to bear a strong resemblance to a broad range of models used in theoretical ecology,
including Lotka-Volterra or predator-prey systems.

We first consider the case where the $N$ languages
in competition in a single area
have equally spaced attractivenesses.
This simple situation allows for an exact characterisation of the stationary state.
The range of attractivenesses is measured by the mean attractiveness $g$.
As this parameter is increased,
the number~$K$ of surviving languages decreases progressively,
as the least favoured languages successively become extinct
at threshold values of $g$.
Importantly, the range of values of $g$
between the start of the disappearances and the appearance of consensus
grows proportionally to $N^2$.
There is therefore a substantial amount of coexistence between languages
that are significantly attractive.

In the general situation,
where the attractivenesses of the competing languages
are modelled as random variables with an arbitrary distribution,
the outcomes of numerical studies at finite $N$ are corroborated
by a detailed asymptotic analysis in the regime of large $N$.
One of the key results is that the quantity $W=Nw$
(the product of the number of languages $N$ with the mean attractiveness~$w$)
determines many quantities of interest,
including the mean fraction $R=\mean{K}/N$ of surviving languages.
The relation between $W$ and $R$ is however non-universal,
as it depends on the full attractiveness distribution.
This non-universality is most prominent in the regime
where the mean attractiveness is large,
so that only the few most favoured languages survive in the stationary state.
The number of such survivors is found to obey a scaling law,
whose non-universal critical exponent is dictated by the specific form
of the attractiveness distribution near its upper edge.

As far as symmetry breaking via spatial heterogeneity is concerned,
we consider the paradigmatic case of two competing languages
in a linear array of $M$ geographic areas, whose neighbours are
linked via a uniform migration rate $\gamma$.
In the simplest situation of two areas,
we determine the full phase diagram of the model
as a function of $\gamma$ as well as the conversion rates
ruling language shift in each area.
This allows us to associate
different regions of phase space with either consensus or coexistence.
Our analysis is then generalised to longer arrays of $M$ linked geographical regions.
We first consider ordered attractiveness profiles,
where language~1 is favoured in the $K$ leftmost areas,
while language~2 is favoured in the $L$ rightmost ones.
If the two blocks are of equal size
so that no language is globally preferred,
coexistence always results; however, the
spatial profiles of the language speakers themselves are rather non-trivial.
For blocks of unequal size,
there is a transition from a situation of coexistence at low migration rates
to a situation of uniform consensus at high migration rates,
where the language favoured in the larger block is the only survivor in all areas.
The critical migration rate at this transition grows as $M^2$.
We next investigate disordered attractiveness profiles,
where conversion rates are modelled as random variables.
There, the probability of observing a uniform consensus
is given by a universal scaling function of $x=\gamma/(M^{3/2}w)$,
where $w$ is the width of the symmetric distribution of conversion rates.

The ratio between migration and conversion rates
beyond which there is consensus
-- either with certainty or with a sizeable probability --
grows with the number of geographic areas
as $M^2$ for ordered profiles of attractivenesses,
and as $M^{3/2}$ for disordered ones.
The first exponent is a consequence of the diffusive nature of migrations,
whereas the second one has been derived in Appendix~\ref{aarrays}
and related to anomalous band-edge scaling in one-dimensional Anderson localisation.
If geographical areas were arranged according to a more complex geometric structure,
these exponents would respectively read $2d/d_s$ and $(4-d_s)/(2d_s)$,
with $d$ and $d_s$ being the fractal and spectral dimensions
of the underlying structure (see~\cite{AO,RT}, and~\cite{HB,HK} for reviews).

Finally, we remark on another striking formal analogy -- that
between the rate equations~(\ref{eq}) presented here,
and those of a spatially extended model of competitive dynamics~\cite{us2},
itself inspired by a model of interacting black holes~\cite{us1}.
In the latter, the non-trivial patterns
of survivors on various networks and other geometrical structures
were a particular focus of investigation,
and led to the unearthing of universal behaviour.
We believe that a network model of competing languages which combines both
the symme\-try-breaking scenarios discussed in this paper,
so that every node corresponds to a geographical area
with its own imbalanced internal dynamics, might lead to the discovery of similar universalities.

\begin{acknowledgement}

AM warmly thanks the Leverhulme Trust for the
Visiting Professorship that funded this research, as well as the Faculty
of Linguistics, Philology and Phonetics at the University of Oxford, for their hospitality.

\noindent
Both authors contributed equally to the present work,
were equally involved in the preparation of the manuscript,
and have read and approved the final manuscript.

\end{acknowledgement}

\appendix

\section{Asymptotic analysis for a large number of competing languages in a single area}
\setcounter{equation}{0}
\def\theequation{A.\arabic{equation}}
\label{appa}

This Appendix is devoted to an analytical investigation
of the statistics of surviving languages in a single geographic area,
in the regime where the numbers $N$ of competing languages is large.

The properties of the attractiveness distribution of the languages are key to
determining whether coexistence or consensus will prevail.
In particular the transition to consensus depends critically, and non-universally, on
the way in which the attractiveness distribution decays, as will be shown below.

Statistical fluctuations between various instances of the model
become negligible for large $N$,
so that sharp (i.e., self-averaging) expressions can be obtained for many
quantities of interest.

Let us begin with the simplest situation where all languages survive.
When the number $N$ of competing languages is large,
the condition for this to occur assumes a simple form.
Consider the expression~(\ref{gxim}) for~$x_N$.
The law of large numbers ensures that the sum $S$ converges to
\beq
W=Nw,
\eeq
whereas $a_N$ is relatively negligible.
The condition that all the $N$ competing languages survive
therefore takes the form of a sharp inequality at large~$N$, i.e.,
\beq
W<1.
\label{wall}
\eeq
All over this regime, the expression for~$x_N$ simplifies~to
\beq
\Lim N x_N=1-W.
\label{xmuall}
\eeq

The above analysis can be extended to the general situation
where the numbers~$N$ of competing languages and~$K$ of surviving ones are
large and comparable, with the fraction of surviving languages,
\beq
R=\frac{K}{N},
\eeq
taking any value in the range $0<R<1$.

The rescaled attractiveness of the least favoured surviving language,
namely
\beq
\eta=\xi_K,
\eeq
turns out to play a key role in the subsequent analysis.
Let us introduce for further reference the truncated moments ($k=0,1,2$)
\beq
I_k(\eta)=\int_\eta^\infty\xi^k\,f(\xi)\,\dd\xi.
\eeq

First of all,
the relationship between $R$ and $\eta$ becomes sharp in the large-$N$ regime.
We have indeed
\beq
R=\prob{\xi>\eta}=1-F(\eta)=I_0(\eta).
\label{reta}
\eeq
The limits of all quantities of interest can be similarly expressed in terms of~$\eta$.
We have for instance
\beq
\Lim S=W\,I_1(\eta),
\eeq
for the sum introduced in~(\ref{s1}).
The marginal stability condition,
namely that language number $K$ is at the verge of becoming extinct,
translates~to
\beq
W=\frac{1}{I_1(\eta)-\eta I_0(\eta)}.
\label{weta}
\eeq
The asymptotic dependence of the fraction $R$ of surviving languages
on the rescaled mean attractiveness $W$
is therefore given in parametric form by~(\ref{reta}) and~(\ref{weta}).
The identity
\beq
\frac{\dd R}{\dd W}=-\frac{f(\eta)}{RW^2}
\eeq
demonstrates that $R$ is a decreasing function of $W$, as it should be.

When the parameter $W$ reaches unity from above,
the model exhibits a continuous transition from the situation where all languages survive.
The parameter $\eta$ vanishes linearly as
\beq
\eta\approx W-1,
\eeq
with unit prefactor, irrespective of the attractiveness distribution.
The fraction of surviving languages departs linearly from unity, according to
\beq
R\approx1-f(0)(W-1).
\eeq

In the regime where $W\gg1$,
the fraction $R$ of surviving languages is expected to fall off to zero.
As a consequence of~(\ref{reta}), $R\ll1$ corresponds to the parameter $\eta$
being close
to the upper edge of the attractiveness distribution $f(\xi)$.
This is to be expected, as the last surviving languages are the most
attractive ones.
As a consequence, the form of the relationship between $W$ and $R$ for $W\gg1$
is highly non-universal,
as it depends on the behavior of the distribution $f(\xi)$ near its upper edge.
It turns out that the following two main classes of attractiveness
distributions have to be considered.

\begin{itemize}

\item
{\it Class~1: Power law at finite distance.}

Consider the situation where the distribution $f(\xi)$ has a finite upper edge $\xi_0$,
and either vanishes or diverges as a power law near this edge, i.e.,
\beq
f(\xi)\approx A\alpha (\xi_0-\xi)^{\alpha-1}.
\eeq
The exponent $\alpha$ is positive.
The density $f(\xi)$ diverges near its upper edge~$\xi_0$ for $0<\alpha<1$,
whereas it vanishes near $\xi_0$ for $\alpha>1$,
and takes a constant value $f(\xi_0)=A$ for $\alpha=1$.

In the relevant regime where $\eta$ is close to $\xi_0$,
the expressions~(\ref{reta}) and~(\ref{weta}) simplify to
\beqa
R&\approx&A\alpha\int_\eta^{\xi_0}(\xi_0-\xi)^{\alpha-1}\dd\xi
\nonumber\\
&\approx&A (\xi_0-\eta)^\alpha,
\\
\frac{1}{W}&\approx&A\alpha\int_\eta^{\xi_0}(\xi-\eta)(\xi_0-\xi)^{\alpha-1}\dd\xi
\nonumber\\
&\approx&\frac{A}{\alpha+1}(\xi_0-\eta)^{\alpha+1}.
\eeqa
Eliminating $\eta$ between both above estimates,
we obtain the following power-law relationship between $W$ and~$R$:
\beq
R\approx\left(\frac{A (\alpha+1)^\alpha}{W^\alpha}\right)^{1/(\alpha+1)}.
\label{rclass1}
\eeq
In terms of the original quantities $K$ and~$w$, the above result reads
\beq
K\approx\left(\frac{A (\alpha+1)^\alpha N}{w^\alpha}\right)^{1/(\alpha+1)}.
\eeq
Setting $K=1$ in this estimate,
we predict that the consensus probability $\P$ becomes appreciable when
\beq
w\sim N^{1/\alpha}.
\label{cclass1}
\eeq

\item
{\it Class~2: Power law at infinity.}

Consider now the situation where the distribution extends up to infinity,
and falls off as a power law, i.e.,
\beq
f(\xi)\approx B\beta\xi^{-\beta-1}.
\eeq
The exponent $\beta$ is larger than 2,
in order for the first two moments of $\xi$ to be convergent.

In the relevant regime where $\eta$ is large,
the expressions~(\ref{reta}) and~(\ref{weta}) simplify to
\beqa
R&\approx& B\beta\int_\eta^\infty\xi^{-\beta-1}\dd\xi
\nonumber\\
&\approx&B\eta^{-\beta},
\\
\frac{1}{W}&\approx& B\beta\int_\eta^\infty\!(\xi-\eta)\xi^{-\beta-1}\dd\xi
\nonumber\\
&\approx&\frac{B}{\beta-1}\eta^{-(\beta-1)}.
\eeqa
Eliminating $\eta$ between both above estimates,
we obtain the following power-law relationship between $W$ and~$R$:
\beq
R\approx\left(\frac{(\beta-1)^\beta}{B W^\beta}\right)^{1/(\beta-1)}.
\label{rclass2}
\eeq
In terms of the original quantities $K$ and~$w$, the above result reads
\beq
K\approx\left(\frac{(\beta-1)^\beta}{B w^\beta N}\right)^{1/(\beta-1)}.
\eeq
Setting $K=1$ in this estimate,
we predict that the consensus probability $\P$ becomes appreciable when
\beq
w\sim N^{-1/\beta}.
\label{cclass2}
\eeq

\end{itemize}

We now summarise the above discussion.
In the regime where $W\gg1$,
the fraction $R$ of surviving languages falls off as a power law of the form
\beq
R\sim\frac{1}{W^\lambda},
\eeq
where the positive exponent $\lambda$ varies continuously,
according to whether the distribution of attractivenesses
extends up to a finite distance or
infinity (see~(\ref{rclass1}),~(\ref{rclass2})):
\beq
\matrix{
\mbox{Class~1:}\quad\hfill&\lambda=\frad{\alpha}{\alpha+1}\quad\hfill&
(\alpha>0,\;0<\lambda<1),\hfill\cr
\mbox{Class~2:}\quad\hfill&\lambda=\frad{\beta}{\beta-1}\quad\hfill&
(\beta>2,\;1<\lambda<2).\hfill
\label{lam}
}
\eeq
In the marginal situation between both classes mentioned above,
comprising e.g.~the exponential distribution,
the decay exponent sticks to its borderline value
\beq
\lambda=1.
\label{lam1}
\eeq
The decay law $R\sim1/W$ might however be affected by logarithmic corrections.

Another view of the above scaling laws goes as follows.
When the number of languages $N$ is large,
the number of surviving languages decreases from $K=N$ to $K=1$
over a very broad range of mean attractivenesses.
The condition for all languages to survive (see~(\ref{wall}))
sets the beginning of this range as
\beq
w_\lo\approx\frac{1}{N}.
\eeq
The occurrence of a sizeable consensus probability $\P$
sets the end of this range~as
\beq
w_\hi\sim N^\mu,
\label{wmu}
\eeq
where the exponent $\mu>-1/2$ varies continuously,
according to (see~(\ref{cclass1}),~(\ref{cclass2})):
\beq
\matrix{
\mbox{Class~1:}\quad\hfill&\mu=\frad{1}{\alpha}\quad\hfill&
(\alpha>0,\;\mu>0),\hfill\cr
\mbox{Class~2:}\quad\hfill&\mu=-\frad{1}{\beta}\quad\hfill&
(\beta>2,\;-1/2<\mu<0).\hfill
\label{mu}
}
\eeq
In the marginal situation between both classes,
the above exponent sticks to its borderline value
\beq
\mu=0.
\label{muc}
\eeq

The extension of the dynamical range,
defined as the ratio between both scales defined above, diverges as
\beq
\frac{w_\hi}{w_\lo}\sim N^{\mu+1}.
\eeq
We predict in particular a linear divergence for the exponential distribution
($\mu=0$)
and a quadratic divergence for the uniform distribution ($\mu=1$).
This explains the qualitative difference observed in Figure~\ref{ave10}.
The slowest growth of the dynamical range is the square-root law
observed for distributions falling off as a power-law
with $\beta\to2$, so that $\mu=-1/2$.

To close, let us underline that most of the quantities met above assume simple forms
for the uniform and exponential distributions (see~(\ref{dis})).

\begin{itemize}

\item
{\it Uniform distribution.}

The consensus probability (see~(\ref{pcdef})) reads
\beq
\P=\left(1-\frac{1}{2w}\right)^N.
\label{puni}
\eeq
For large $N$, this becomes $\P\approx\exp(-N/(2w))$,
namely a function of the ratio $w/N$,
in agreement with~(\ref{wmu}) and~(\ref{mu}), with exponent $\mu=1$, since $\alpha=1$.

The truncated moments read
\beqa
I_0(\eta)&=&1-\frac{\eta}{2},\quad
I_1(\eta)=1-\frac{\eta^2}{4}.
\eeqa
We thus obtain
\beq
R=\frac{1}{\sqrt W},
\label{runi}
\eeq
with exponent $\lambda=1/2$, in agreement with~(\ref{rclass1})
and~(\ref{lam}) for $\alpha=1$.

\item
{\it Exponential distribution.}

The consensus probability reads
\beq
\P=\e^{-1/w},
\label{pexp}
\eeq
irrespective of $N$, in agreement with~(\ref{wmu}),
with exponent $\mu=0$ (see~(\ref{muc})).

The truncated moments read
\beqa
I_0(\eta)&=&\e^{-\eta},\quad
I_1(\eta)=(1+\eta)\e^{-\eta}.
\eeqa
We thus obtain
\beq
R=\frac{1}{W},
\label{rexp}
\eeq
with exponent $\lambda=1$, in agreement with~(\ref{lam1}).

\end{itemize}

\section{Stability matrices and their spectra}
\setcounter{equation}{0}
\def\theequation{B.\arabic{equation}}
\label{appb}

\subsection{Generalities}
\label{agal}

This Appendix is devoted to stability matrices and their spectra.
Let us begin by reviewing some general background
(see e.g.~\cite{KH} for a comprehensive overview).
Consider an autonomous dynamical system
defined by a vector field $\m E(\m x)$ in $N$ dimensions,
i.e., by~$N$ coupled first-order equations of the form
\beq
\der{x_m(t)}=E_m\{x_n(t)\},
\eeq
with $m,n=1,\dots,N$,
where the right-hand sides depend on the dynamical variables $\{x_n(t)\}$
themselves,
but not explicitly on time.

Assume the above dynamical system has a fixed point $\{x_m\}$,
such that $E_m\{x_n\}=0$ for all~$m$.
Small deviations $\{\delta x_m(t)\}$ around the fixed point $\{x_m\}$
obey the linearised dynamics given by the stability matrix $\m S$,
i.e., the $N\times N$ matrix defined by
\beq
S_{m,n}=\frac{\partial E_m}{\partial x_n},
\eeq
where right-hand sides are evaluated at the fixed point.
The fixed point is stable,
in the strong sense that small deviations fall off exponentially fast to zero,
if all eigenvalues $\lambda_a$ of $\m S$ have negative real parts.
In this case, if all the~$\lambda_a$ are real,
their opposites $\omega_a=-\lambda_a>0$ are the inverse relaxation times
of the linearised dynamics.
In particular, the opposite of the smallest eigenvalue, simply denoted by
$\omega$,
characterises exponential convergence to the fixed point for a generic initial state.
If some of the $\lambda_a$ have non-zero imaginary parts, convergence is
oscillatory.

The analysis of fixed points and bifurcations in low-dimensional
Lotka-Volterra and replicator equations
has been the subject of extensive investigations~\cite{hofna,tj,bom1,s+s1,bom2,s+s2}
(see also~\cite{hofbook,hofbams,nsscience}).

\subsection{Array models}
\label{aarrays}

The remainder of this Appendix is devoted to the stability matrices
involved in the array models considered in Section~\ref{spatial},
for an arbitrarily large number $M$ of geographical areas.
All those stability matrices are related to the symmetric $M\times M$ matrix
\beq
\m\Delta_M=\pmatrix{
1 & -1 & 0 &\dots\cr
-1 & 2 & -1 &\dots\cr
\dots &\dots &\dots &\dots\cr
\dots & -1 & 2 & -1\cr
\dots & 0 & -1 & 1},
\label{lap}
\eeq
representing (minus) the Laplacian operator
on a linear array of $M$ sites, with Neumann boundary conditions.
References~\cite{wilson,bollobas} provide reviews
on the Laplacian and related operators on graphs.

The eigenvalues $\lambda_a$ of $\m\Delta_M$
and the corresponding normalised eigenvectors~$\m\phi_a$,
such that $\m\Delta_M\m\phi_a=\lambda_a\m\phi_a$
and $\m\phi_a\cdot\m\phi_b=\delta_{ab}$, read
\beqa
\lambda_a&=&4\sin^2\frac{a\pi}{2M},
\nonumber\\
\phi_{a,m}&=&\sqrt\frac{2}{(1+\delta_{a0})M}\,\cos\frac{(2m-1)a\pi}{2M}
\label{deltaspec}
\eeqa
($a=0,\dots,M-1$).
The vanishing eigenvalue $\lambda_0=0$ corresponds to the uniform eigenvector
$\phi_{0,m}=1/\sqrt{M}$.

Let us begin by briefly considering the simple example of the stability matrix
\beq
\m S_M=-\m 1-\gamma\m\Delta_M.
\label{stap}
\eeq
of the rate equations~(\ref{ap}) for the total populations $P_m(t)$.
Its eigenvalues are $-1-\gamma\lambda_a$.
The smallest of them is $-1$, so that the inverse relaxation time
is given by $\omega=1$, as announced below~(\ref{ap}).

Let us now consider the stability matrices
\beqa
&&\m S_M^\un=-\diag(C_1,\dots,C_M)-\gamma\m\Delta_M,
\nonumber\\
&&\m S_M^\ze=\diag(C_1,\dots,C_M)-\gamma\m\Delta_M.
\eeqa
respectively defined in~(\ref{sun}) and~(\ref{sze}),
and corresponding to both uniform consensus states
for an arbitrary profile of conversion rates $C_m$.
The ensuing stability conditions have been written down explicitly
in~(\ref{cdun}) and~(\ref{cdze}) for $M=2$.
It will soon become clear that it is virtually impossible to write them down
for an arbitrary size $M$.
Some information can however be gained from the calculation of the determinants
of the above matrices.
They only differ by a global sign change of all the conversion rates $C_m$,
so that it is sufficient to consider $\m S_M^\un$.
It is a simple matter to realise that its determinant reads
\beq
D_M=\det\m S_M^\un=(-\gamma)^M(u_{M+1}-u_M),
\label{detres}
\eeq
where $u_m$ is a generalised eigenvector solving the following Cauchy problem:
\beq
-(C_m+2\gamma)u_m+\gamma(u_{m+1}+u_{m-1})=0,
\label{cau}
\eeq
with initial conditions $u_0=u_1=1$.
We thus obtain recursively
\beqa
\gamma u_2&=&C_1+\gamma,
\nonumber\\
-D_1&=&C_1,
\\
\gamma^2u_3&=&C_1C_2+\gamma(2C_1+C_2)+\gamma^2,
\nonumber\\
D_2&=&C_1C_2+\gamma(C_1+C_2),
\label{d2res}
\\
\gamma^3u_4&=&C_1C_2C_3+\gamma(2C_1C_2+2C_1C_3+C_2C_3)
\nonumber\\
&+&\gamma^2(3C_1+2C_2+C_3)+\gamma^3,
\nonumber\\
-D_3&=&C_1C_2C_3+\gamma(C_1C_2+2C_1C_3+C_2C_3)
\nonumber\\
&+&\gamma^2(C_1+C_2+C_3),
\label{d3res}
\eeqa
and so on.
The expression~(\ref{d2res}) for $D_2$ agrees
with the second of the conditions~(\ref{cdun})
and with the equation of the red curve in Figure~\ref{c1c2plot}, as should be.
The expression~(\ref{d3res}) for $D_3$ demonstrates that the complexity of
the stability conditions grows rapidly with the system size $M$.

\subsubsection*{Random arrays}

In the case of random arrays, considered in Section~\ref{srandom},
the conversion rates $C_m$ are independent random variables
such that $\mean{C_m}=0$ and $\mean{C_m^2}=w^2$.

The regime of most interest is where the conversion rates $C_n$
are small with respect to $\gamma$.
In this regime,
the determinant $D_M$ can be expanded as a power series in the conversion rates.
The $u_m$ solving the Cauchy problem~(\ref{cau}) are close to unity.
Setting
\beq
u_m=1+u_m^\un+u_m^\de+\cdots,
\eeq
where the $u_m^\un$ are linear and the $u_m^\de$ quadratic in the $C_n$,
we obtain after some algebra
\beq
(-)^MD_M=\gamma^{M-1}(X+Y+\dots),
\label{dsmall}
\eeq
where
\beq
X=\sum_{n=1}^MC_n,\quad
Y=\frac{1}{2\gamma}\sum_{m,n=1}^M\abs{n-m}C_mC_n
\eeq
are respectively linear and quadratic in the $C_n$.
We have
\beqa
&&\mean{X}=\mean{Y}=\mean{XY}=0,
\nonumber\\
&&\sigma_X^2=\mean{X^2}=Mw^2,
\nonumber\\
&&\sigma_Y^2=\mean{Y^2}=\frac{M^2(M^2-1)w^4}{12\gamma^2}.
\eeqa

In Section~\ref{srandom}
we need an estimate of the probability~$\Q$ that $\overline C=X/M$ is atypically small.
Within the present setting, it is natural to define the latter event as $\abs{X}<\abs{Y}$.
The corresponding probability can be worked out
proviso we make the ad hoc simplifying assumptions
-- that definitely do not hold in the real world --
that $X$ and $Y$ are Gaussian and independent.
Within this framework,
the complex Gaussian random variable
\beq
\zeta=\frac{X}{\sigma_X}+\frac{\ii Y}{\sigma_Y}
\eeq
has an isotropic density in the complex plane.
We thus obtain
\beq
\Q=\frac{2}{\pi}\arctan\frac{\sigma_Y}{\sigma_X}
\approx\frac{2}{\pi}\arctan\frac{M^{3/2}w}{\gamma\sqrt{12}}.
\label{qrough}
\eeq

\subsubsection*{Ordered arrays}

The aim of this last section is to investigate the spectrum
of the stability matrix $\m S_M^\un$
associated with the ordered profile of conversion rates given by~(\ref{cord}).

In this case, the generalised eigenvector $u_m$ solving the Cauchy
problem~(\ref{cau})
can be worked out explicitly.
We have $C_m=1$ for $m=1,\dots,K$,
and therefore $u_m=a\e^{m\mu}+b\e^{-m\mu}$,
where $\mu>0$ obeys the dispersion relation
\beq
4\gamma\sinh^2\muh=1.
\label{imagdisp}
\eeq
The initial conditions $u_0=u_1=1$ fix $a$ and $b$, and so
\beq
u_m=\frac{\cosh(2m-1)\muh}{\cosh\muh}\quad(m=0,\dots,K+1).
\eeq
Similarly, we have $C_m=-1$ for $m=K+\ell$, with $\ell=1,\dots,L$,
and therefore $u_m=\alpha\e^{\ii\ell q}+\beta\e^{-\ii\ell q}$,
where $0<q<\pi$ obeys the dispersion relation
\beq
4\gamma\sin^2\qh=1.
\eeq
Matching both solutions for $m=K$ and $K+1$ fixes $\alpha$ and~$\beta$, and so
\beqa
u_m&=&\frac{\cosh(2K+1)\muh\sin\ell
q-\cosh(2K-1)\muh\sin(\ell-1)q}{\cosh\muh\sin q}
\nonumber\\
&&(m=K+\ell;\;\ell=0,\dots,L+1).
\eeqa
Inserting the latter result into~(\ref{detres}),
we obtain the following expression for the determinant of $\m S_M^\un$, with
$M=K+L$:
\beqa
D_M&=&2(-\gamma)^M
\\
&\times&\left(\tanh\muh\sinh K\mu\cos Lq-\tan\qh\cosh K\mu\sin Lq\right).
\nonumber
\eeqa
The vanishing of the above expression, i.e.,
\beq
\tanh\muh\tanh K\mu=\tan\qh\tan Lq,
\label{tata}
\eeq
signals that one eigenvalue of the stability matrix $\m S^\un$ vanishes.
In particular, the consensus state where language~1 survives
becomes marginally stable at the threshold migration rate~$\gamma_c$,
where the largest eigenvalue of $\m S^\un$ vanishes.
Equation~(\ref{tata}) amounts to a polynomial equation of the form $P_{K,L}(\gamma)=0$,
where the polynomial $P_{K,L}$ has degree $K+L-1=M-1$.
All its zeros are real, and~$\gamma_c$ is the largest of them.
The first of these polynomials read
\beqa
P_{2,1}&=&\gamma^2-2\gamma-1,
\nonumber\\
P_{3,1}&=&2\gamma^3-2\gamma^2-4\gamma-1,
\nonumber\\
P_{4,1}&=&3\gamma^4-9\gamma^2-6\gamma-1,
\nonumber\\
P_{3,2}&=&\gamma^4-10\gamma^3-7\gamma^2+2\gamma+1.
\eeqa

\bibliography{revised.bib}

\end{document}